\documentclass[showpacs,preprintnumbers,amsmath,amssymb]{revtex4}
\usepackage{graphicx}
\begin{document}
%\baselineskip 20.75pt
 %=========================================================================

%\vspace{0.5cm} \preprint{{\em Submitted to Phys.~Rev.~Lett.} %\hspace{3.0in} Web galley}
\date{\today}
\vspace{2.7in}

\title{Drag effects in the system of electrons and microcavity polaritons}

\author{Oleg L. Berman$^{1}$, Roman Ya. Kezerashvili$^{1,2}$, and Yurii E.
Lozovik$^{3,4}$} \affiliation{\mbox{$^{1}$Physics Department, New York
City College of Technology, The City University of New York,} \\
Brooklyn, NY 11201, USA \\
\mbox{$^{2}$The Graduate School and University Center, The City
University of New York,}
New York, NY 10016, USA \\
\mbox{$^{3}$Institute of Spectroscopy, Russian Academy of
Sciences,} \\
142190 Troitsk, Moscow Region, Russia \\
\mbox{$^{4}$Moscow Institute of Physics and Technology
(State University), 141700, Dolgoprudny, Russia }}

\begin{abstract}

The theory of the drag effects in the system of spatially separated electrons and excitons in
coupled quantum wells (QW) embedded in an optical microcavity is developed. It is shown that at
low temperature an electron current induces the (normal component) polariton flow, therefore, a
transport of photons along the cavity. However, the electron current dragged by the polariton
flow is strongly suppressed below polariton superfluid transition temperature and hence, the
strong suppression of the induced electron current indicates the superfluidity of polaritons.
Therefore, the transport properties of polaritons can be investigated by measuring the current or
voltage in the electron subsystem. At high temperatures we study the exciton-electron drag
effects. At high temperatures regime, from one hand, the existence of the electric current in an
electron QW induces the exciton flow in the other QW, from the other hand, the electron current
in one QW induces the exciton flow in the other QW via the drag of excitons by the electrons. The
drag coefficients for the polariton-electron systems are calculated and analyzed. We discuss the
possible experimental observation of the drag effects in the system of electrons and microcavity
polaritons, that also allow to observe the cavity polaritons superfluidity.

\vspace{0.1cm}

\end{abstract}


 \maketitle
%\newpage

%-------------------------------------------------------------------------
%-------------------------------------------------------------------------

\section{Introduction}
\label{int}

The study of the drag effects in a two-layer system of electrons and
holes, and electrons and excitons has long history. The drag effects
in a two-layer system of spatially separated electrons and holes
were predicted theoretically, and their influence on phase
transitions to a superfluid excitonic phase investigated in
Ref.~[\onlinecite{Lozovik}].  Later, Pogrebinskii in
Ref.~[\onlinecite{Pogrebinskii}] discussed the drag of electrons by
electrons in a semiconductor$-–$insulator$-$semiconductor structure.
Price proposed a practical method for observing the drag effect in
heterostructures~\cite{Price}. Subsequently, the drag effect was
explored in a number of theoretical and experimental
studies~\cite{Gramila,Sivan,Gramila2,Jauho,Zheng,Sirenko,Tso,Flensberg,Vignale,Tanatar},
where various physical realizations of the drag effect were
discussed in one-dimensional, two-dimensional, and three-dimensional
systems. The prediction was that for two conducting layers separated
by an insulator there will be a drag of carriers in one film due to
the direct Coulomb interaction with the carriers in the other film.

A theory of transresistivity coefficients, which relate electric
fields in one layer to currents flowing in the opposite layer, for
the case of an electron-hole double-layer system with a superfluid
electron-hole condensate  was presented in
Ref.~[\onlinecite{Vignale}]. Such electron-hole system can be
realized in GaAs/AlGaAs coupled quantum wells (CQWs). The
measurement of these transport coefficients could provide an
unambiguous experimental indication of the existence of the
condensate. When no condensate is present, the transresistivity is
typically several orders of magnitude smaller than the isolated
layer resistivity. According to Ref.~[\onlinecite{Vignale}], the
transresistivity jumps to a value comparable to the isolated layer
resistivity as soon as the condensate forms, it continues to
increase with decreasing temperature $T$, and it diverges as $T
\rightarrow 0$. Employing diagrammatic perturbation theory, the
charge  Coulomb drag resistivity and spin Coulomb drag resistivity
were calculated in the presence of Rashba spin-orbit coupling in
Ref.~[\onlinecite{Das_Sarma_spin}].

For the past decade there was great amount of various interesting experiments on drag effects
performed in the different groups. The measurements in a perpendicular magnetic field of the
frictional drag between two closely spaced, but electrically isolated, two-dimensional electron
gases were presented by Pepper's group~\cite{Pepper}.  The drag measurements on dilute double
layer two-dimensional hole systems have been performed in  Tsui's group~\cite{Tsui}. The
formation of the superfluid exciton Bose condensate at low temperatures has been experimentally
studied by performing magnetotransport and drag measurements on a quasi-Corbino two-dimensional
electron bilayer at a total filling factor of 1 by  von Klitzing's group~\cite{von_Klitzing}.
Electron-hole bilayers are expected to make a transition from a pair of weakly coupled
two-dimensional systems to a strongly coupled exciton system as the barrier between the layers is
reduced. The recent measurements done by Lilly's group~\cite{Lilly2} for Coulomb drag in the
exciton regime in electron-hole bilayers demonstrated that there is an increase in the drag
resistance as the temperature is reduced when a current is driven in the electron layer and
voltage measured in the hole layer. These results indicate the onset of strong coupling possibly
due to exciton formation or phenomena related to exciton condensation.

The kinetic properties of a two-layer system of electrons and excitons at temperatures above the
temperature  $T_{c}$  of the Kosterlitz--Thouless transition \cite{Kosterlitz}, at which there is
no local condensate or superfluidity in the exciton system, were considered in
Ref.~[\onlinecite{Nikitkov1}]. The kinetic properties of a two-layer system of electrons and
excitons at temperatures below the temperature  $T_{c}$ under conditions allowing the existence
of a Bose condensate of excitons have been studied in Ref.~[\onlinecite{Nikitkov2}]. The
corresponding calculations have been performed for two-dimensional systems of spatially separated
electrons and excitons. The effect of spatially separated electron$–-$exciton drag in a double
layer system was studied in the Debye--H\"{u}ckel approximation taking into account screening of
the interlayer electron$–-$exciton interaction~\cite{Kulakovskii}.

Recently   Bose coherent effects of two-dimensional (2D) excitonic
polaritons in a quantum well embedded in a optical microcavity have
been the subject of theoretical and experimental studies
\cite{pssb,book,Snoke_text}. To obtain polaritons, two mirrors
placed opposite each other form a microcavity, and quantum wells are
embedded within the cavity at the antinodes of the confined optical
mode. The resonant exciton-photon interaction results in the Rabi
splitting of the excitation spectrum.  Two polariton branches appear
in the spectrum due to the resonant exciton-photon coupling. The
lower polariton
  branch of the spectrum has a minimum at zero momentum.
 The effective mass of the lower polariton is extremely  small, and lies in the range $10^{-5}-10^{-4}$ of the free electron mass.
  These lower polaritons form a 2D weakly interacting Bose gas. The extremely light mass of these bosonic quasiparticles at experimentally
  achievable excitonic  densities, results in a
relatively high critical  temperature for superfluidity, of $100 \
\mathrm{K}$ or even higher, because  the 2D thermal de Broglie
wavelength is inversely proportional
 to the mass of the quasiparticle.

While at finite temperatures there is no true BEC  in any infinite
untrapped 2D system, a true 2D BEC quantum phase transition can be
obtained in the presence of a confining potential
\cite{Bagnato,Nozieres}. The essential experimental progress was
achieved in experimental studies of exciton polaritons in the system
of a QW embedded in optical microcavity
\cite{Dang,Yamamoto,Baumberg}.
  Recently,  the  polaritons in a harmonic potential trap have  been studied experimentally in a GaAs/AlAs
   quantum well embedded in a GaAs/AlGaAs microcavity \cite{Balili}. In this trap,
   the exciton energy is shifted using a stress-induced
band-gap. In this system, evidence for the BEC of polaritons in a
quantum well has been observed \cite{science,Balili_prb}. The theory
of the BEC and superfluidity of excitonic polaritons in a quantum
well in a parabolic trap has been developed in
Ref.~[\onlinecite{Berman_L_S}]. The Bose condensation of polaritons
is caused by their bosonic character
\cite{science,Berman_L_S,Kasprzak}. The BEC and superfluidity of
cavity polaritons in a QW without a trap were considered
in~[\onlinecite{Kavokin2,Kavokin3,Lit,Amo1,Amo2}].

The investigation of  the kinetic properties of a system of spatially separated polaritons and
electrons  based on drag effects can provide additional information regarding the phase state of
the polariton subsystem and phase transitions in it. The phase state of the polariton subsystem
can be analyzed by performing a simple study of the response of the electron subsystem. In other
words, the transport properties of polaritons and their changes upon phase transitions can be
investigated by measuring the current or voltage in the electron subsystem. Another property of
systems of spatially separated interacting quasiparticles
 is the possibility of controlling the motion
of the quasiparticles of one subsystem by altering the parameters of
state of the quasiparticles in the other subsystem (for example,
controlling the motion of electrons using a flow of polaritons).
While the drag effect in various electron$-$hole systems were
considered in many interesting papers cited above, we are lacked of
such research for a  polariton-electron system. The drag effect in
the polariton-electron system embedded in an optical microcavity was
predicted  just recently~\cite{BKL_drag}. In this Paper we develop
the theory of the drag effects in the system of spatially separated
electrons and excitons in coupled quantum wells embedded in an
optical microcavity.

The Paper is organized in the following way. In Sec.~\ref{inter} we
present the interaction Hamiltonian between the spatially separated
polaritons and electrons. In Sec.~\ref{rel} we study transport
relaxation time of the quasiparticle excitations and polaritons. In
Sec.~\ref{drag_pe} we calculate the drag coefficients corresponding
to the drag of the quasiparticles in the polariton subsystem by the
electron current and  the drag of the electrons by the
quasiparticles in the polariton subsystem. The study of the drag
effects in the exciton-electron system at high temperatures is
presented in Sec.~\ref{drag_exe}. In Sec.~\ref{experiment} we
propose the experiments to observe the drag effects.  Finally, the
discussion of the results and conclusions follow in Sec.~\ref{disc}.

%-------------------------------------------------------------------------
%-------------------------------------------------------------------------
\section{Interaction of polaritons with electrons}
\label{inter}

We consider two neighboring quantum wells embedded in an optical
microcavity: the first QW is occupied by 2D electron gas (2DEG) and
the second QW is occupied by the excitons created by the laser
pumping.

At low temperature  $k_{B}T \ll \hbar\Omega_{R}$, where $\hbar \Omega_{R}$ is Rabi splitting, and
$k_{B}$ is Boltzmann constant, and at the resonance of excitons with cavity photons, the excitons
are entangled with the cavity photons and form the exciton polaritons. Here we omit weak effects
of direct nonresonant interaction of photons with 2D electron gas which will be considered
elsewhere. By focusing laser pumping in some region of the cavity the gradient of excitons and
exciton polaritons densities can be generated. These gradients induce the polariton and exciton
flows, and in a turn the normal component of moving excitons drags the electrons in the
neighboring QW due to the electron-exciton interaction. So the electric current would be
generated by the flow of the normal component of exciton polaritons.

In another scenario by applying electric voltage in the QW with 2DEG the electronic current is
induced, and this current drags the normal component of excitons in the  neighboring QW. The
excitons are entangled with the cavity photons. So the cavity photons can be also dragged and,
thus, controlled by the electric voltage.
 Besides, possible applications of the control of photons and (or) excitons by the exciton-electron drag can be used for study of the properties and phases
 in the polariton and exciton system, particularly, the superfluidity of the system.

\begin{figure}[t] %  figure placement: here, top, bottom, or page
   \centering
  \includegraphics[width=3.5in]{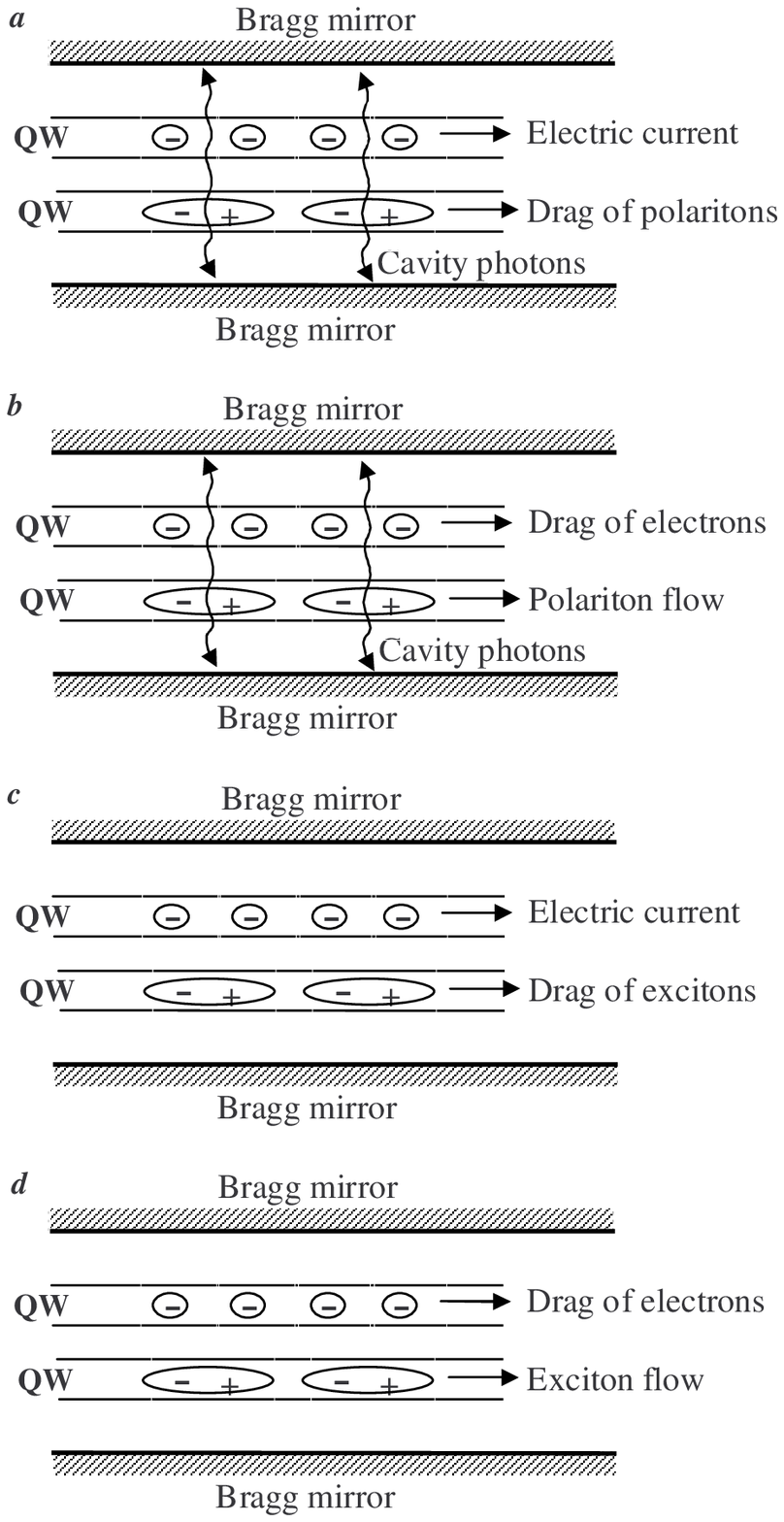}
\caption{The schematic diagram for the drag effects in the CQWs
embedded in an optical microcavity. a. The quasiparticles in the
cavity polariton subsystem are dragged by the electron current at
low temperatures. b. The electrons are dragged by the flow of the
quasiparticles in the cavity polariton subsystem at low
temperatures. c. The excitons are dragged by the electron current at
the high temperatures. d. The electrons are dragged by the exciton
flow at high temperatures.} \label{dr}
\end{figure}

 From the other hand, at high temperature, $k_{B}T \gtrsim \hbar\Omega_{R}$, the polariton states are occupied mainly far from the polariton resonance, and in these states
 exciton-photon entanglement is negligible. Thus, at high temperatures only the exciton-electron drag is
 essential, and the exciton flow can induce the electron current as well as the electron current can produce  the exciton
flow. The various drag effects in an optical microcavity considered
in  this Paper are schematically shown in Fig.~\ref{dr}.

We consider the system of two parallel quantum wells embedded into a semiconductor microcavity.
 One quantum well is filled by two-dimensional exciton polaritons formed by direct excitons and microcavity photons.
 We assume that the exciton system is dilute and weakly interacting system, satisfying to the following condition $n_{ex} a_{0}^{2} \ll 1$, where $n_{ex}$ is the 2D exciton density,
  $a_{0} = \epsilon \hbar^{2}/(2\mu_{0}e^{2})$ is the 2D exciton Bohr radius, $\mu_{0} = m_{e}m_{h}/(m_{e} + m_{h})$ is the reduced exciton mass,
   $m_{e}$ and $m_{h}$ are effective masses of an electron and a hole, respectively, $\epsilon$ is the dielectric constant in a quantum well,
    $e$ is the charge of an electron. The  condition $n_{ex} a_{0}^{2} \ll 1$ holds for the excitons at the exciton densities
    up to $n_{ex} \approx 10^{12} \ \mathrm{cm}^{-2}$, since in GaAs/GaAsAl quantum well the exciton Bohr radius is in the order
     of $a_{0} \sim 10-50 \ \mathrm{{\AA}}$.  It is obvious to conclude that the system of polaritons is also weakly interacting system, since the system of excitons,
     forming the polaritons, is dilute. The other quantum well parallel to the quantum well with excitons is filled by the 2D electron gas (2DEG).
     The polariton-electron drag effects are caused by the
     polariton-electron interaction.

Let us mention that the second QW filled by the 2DEG can also
possess excitons, because the excitonic population can be driven by
the microcavity photons. Therefore, in this system of the coupled
QWs embedded in an optical microcavity, the formation of polaritons,
whose exciton components are shared on both the QWs, can occur.
However, we do not consider the excitons formation in the QW filled
by the 2DEG, since we suggest constructing this 2DEG QW so that the
excitons in this QW are not in the resonance with the microcavity
photons and the excitons in the other, ``excitonic'' QW. This can be
achieved by the fact that the two QWs embedded in an optical
microcavity are assumed to have different chemical compositions or
different widths.

The potential energy of the pair attraction between the exciton and electron placed in two parallel quantum wells with the spatial separation $D$ is given by \cite{Nikitkov1,Nikitkov2}
%%%%%%%%%%%%%%%%%%%%%%%%%%%%%%%%%%%%%%%%%%%%%%%%%%%%%%%%%%%%%%%%%%%%%%%%%%%%%%%%%%%%%%%%%%%%%%%%%%%%%%%%%%%%%%%%%%%%%%%%%%%%%%%%%%%%%%%%%%%%%%%%%%%%%%%%%%%
\begin{eqnarray}\label{ex-el}
W(r,D) = - \frac{21}{32} \frac{e^{2} a_{0}^{3}}{\epsilon (r^{2} + D^{2})^{2}} \  ,
\end{eqnarray}
%%%%%%%%%%%%%%%%%%%%%%%%%%%%%%%%%%%%%%%%%%%%%%%%%%%%%%%%%%%%%%%%%%%%%%%%%%%%%%%%%%%%%%%%%%%%%%%%%%%%%%%%%%%%%%%%%%%%%%%%%%%%%%%%%%%%%%%%%%%%%%%%%%%%%%%%%%%%
where $r$ is the distance between the exciton and electron along the plane.
The 2D Fourier image of $W(r,D)$ is
%%%%%%%%%%%%%%%%%%%%%%%%%%%%%%%%%%%%%%%%%%%%%%%%%%%%%%%%%%%%%%%%%%%%%%%%%%%%%%%%%%%%%%%%%%%%%%%%%%%%%%%%%%%%%%%%%%%%%%%%%%%%%%%%%%%%%%%%%%%%%%%%%%%%%%%%%%%
\begin{eqnarray}\label{ex-el2}
W(q,D) = - \frac{21 \pi }{32} \frac{e^{2} a_{0}^{3}}{\epsilon}
\frac{q}{\hbar D} K_{1}(qD/\hbar) \  ,
\end{eqnarray}
%%%%%%%%%%%%%%%%%%%%%%%%%%%%%%%%%%%%%%%%%%%%%%%%%%%%%%%%%%%%%%%%%%%%%%%%%%%%%%%%%%%%%%%%%%%%%%%%%%%%%%%%%%%%%%%%%%%%%%%%%%%%%%%%%%%%%%%%%%%%%%%%%%%%%%%%%%%%
where $K_{1}(qD)$ is the modified Bessel function of the second kind.

In a many-particle system of excitons and electrons the bare pair interaction $W(q,D)$ should be replaced by the effective interaction  $W_{eff}(q,D)$ corresponding to the exciton-electron interaction screened by the electron-electron interactions in 2DEG. In the system of dilute excitons in one quantum well and 2DEG in the other quantum well $W_{eff}(q,D)$ is given by \cite{Nikitkov2}
%%%%%%%%%%%%%%%%%%%%%%%%%%%%%%%%%%%%%%%%%%%%%%%%%%%%%%%%%%%%%%%%%%%%%%%%%%%%%%%%%%%%%%%%%%%%%%%%%%%%%%%%%%%%%%%%%%%%%%%%%%%%%%%%%%%%%%%%%%%%%%%%%%%%%%%%%%%
\begin{eqnarray}\label{ex-el3}
W_{eff}(q,D) = - \frac{W(q,D)}{1 - \Pi_{e}(q)V_{e}(q)} \  ,
\end{eqnarray}
%%%%%%%%%%%%%%%%%%%%%%%%%%%%%%%%%%%%%%%%%%%%%%%%%%%%%%%%%%%%%%%%%%%%%%%%%%%%%%%%%%%%%%%%%%%%%%%%%%%%%%%%%%%%%%%%%%%%%%%%%%%%%%%%%%%%%%%%%%%%%%%%%%%%%%%%%%%%
where $V_{e}(q) = 2 \pi e^{2}\hbar/(\epsilon q)$ is the 2D Fourier
image of the pair electron-electron Coulomb attraction,
 and $\Pi_{e}(q)$ is the 2D polarization function for electrons. We consider the limit of the very dense 2DEG,
 when the following condition is valid: $n_{el} a_{e}^{2} \gg 1$, where $n_{el}$ is the 2D density of the electrons and
  $a_{e} = \epsilon \hbar^{2}/(2m_{e}e^{2})$ is the 2D electron Bohr radius. In the limit of very dense 2DEG in the Thomas-Fermi approximation
  $\Pi_{e}(q)$ is  $\Pi_{e}(q) = m_{e}/(\pi\hbar^{2})$~\cite{AFS}, and, therefore, $\Pi_{e}(q)V_{e}(q) = (a_{e}q/\hbar)^{-1}$.

At low exciton density,  $n_{ex} a_{0}^{2} \ll 1$, the term of the Hamiltonian $\hat{H}_{ex-el}$ corresponding to the exciton-electron interaction has the form \cite{Nikitkov2}
%-------------------------------------------------------------------------
\begin{eqnarray}
\label{Ham_exc} \hat{H}_{ex-el} = \frac{1}{A}\sum_{\mathbf{p}_{1},\mathbf{p}_{2},\mathbf{p}_{1}^{\prime},\mathbf{p}_{2}^{\prime}} W_{eff}(|\mathbf{p}_{1} - \mathbf{p}_{1}^{\prime}|,D)
\hat{a}_{\mathbf{p}_{1}^{\prime}}^{\dagger}\hat{c}_{\mathbf{p}_{2}^{\prime}}^{\dagger}\hat{c}_{\mathbf{p}_{2}} \hat{a}_{\mathbf{p}_{1}} \ ,
\end{eqnarray}
%-------------------------------------------------------------------------
where $\hat{a}_{\mathbf{p}}^{\dagger}$ and $\hat{a}_{\mathbf{p}}$ are the exciton Bose creation
and annihilation operators, respectively, $\hat{c}_{\mathbf{p}}^{\dagger}$ and
$\hat{c}_{\mathbf{p}}$ are the electron Fermi creation and annihilation operators, respectively,
and $A$ is the macroscopic quantization area of the system. Below for the derivation of the drag
coefficients we need the Hamiltonian of the exciton-electron interaction in the representation of
the operators of the quasiparticles in the polariton subsystem. The details for the derivation of
the Hamiltonian of the exciton-electron interaction in the representation of the quasiparticle
operators are presented in Appendix~\ref{ap.qp}.

%-------------------------------------------------------------------------
%-------------------------------------------------------------------------

\section{Transport relaxation time of the quasiparticle excitations and polaritons}
\label{rel}

In order to calculate polariton-electron and exciton-electron drag
coefficients we use the kinetic equations in Sec.~\ref{drag_pe},
which include the transport relaxation time $\tau_{1}(p)$ of the
quasiparticle excitations  corresponding to the scattering on the
impurities enters. In this Section we obtain the transport
relaxation time  to the scattering of the quasiparticles on the
impurities. The Hamiltonian of elastic interactions of excitons with
impurities is given by
%-------------------------------------------------------------------------
\begin{eqnarray}
\label{imp} \hat{H}_{1} =
\frac{1}{A}\sum_{\mathbf{p},\mathbf{p}^{\prime}}
V(\mathbf{p},\mathbf{p}^{\prime}) \hat{a}_{\mathbf{p}}^{\dagger}
\hat{a}_{\mathbf{p}^{\prime}}  \ ,
\end{eqnarray}
%-------------------------------------------------------------------------
where $V(\mathbf{p},\mathbf{p}^{\prime})$ is is the matrix element for the interactions of an exciton with impurities.
Replacing the exciton operators by the polariton operators according to Eqs.~(\ref{bog_tr}) and the polariton operators
by the operators for quasiparticle excitations defined by Eqs.~(\ref{bog_uv}), and keeping only the terms which satisfy the requirement of elastic collisions, we obtain
%-------------------------------------------------------------------------
\begin{eqnarray}
\label{imp2} \hat{H}_{1} =
\frac{1}{A}\sum_{\mathbf{p},\mathbf{p}^{\prime}}
V(\mathbf{p},\mathbf{p}^{\prime}) \sigma (p,p^{\prime})
\hat{b}_{\mathbf{p}}^{\dagger} \hat{b}_{\mathbf{p}^{\prime}}  \ ,
\end{eqnarray}
%-------------------------------------------------------------------------
Using Hamiltonian (\ref{imp2})  and applying Fermi's golden rule, we
obtain for the reciprocal of the transport relaxation time
$\tau_{1}(p)$
%-------------------------------------------------------------------------
\begin{eqnarray}
\label{t1} \frac{1}{\tau_{1}(p)} = \frac{2\pi}{\hbar} \int  \left|
V(\mathbf{p},\mathbf{p}^{\prime}) \sigma (p,p^{\prime})\right|^{2}
\delta(\varepsilon_{1}(p) - \varepsilon_{1}(p^{\prime})) \left(1 -
\cos(\hat{(\mathbf{p},\mathbf{p}^{\prime})}) \right) \frac{s
d^{2}p^{\prime}}{(2\pi\hbar)^{2}} \ .
\end{eqnarray}
%-------------------------------------------------------------------------

Substituting Eqs.~(\ref{sig}) and~(\ref{bog_uv}) into Eq.~(\ref{t1}), we get
%-------------------------------------------------------------------------
\begin{eqnarray}
\label{t12}
\tau_{1}(p) = \frac{\varepsilon_{1}(p)}{\xi(p)} \tau_{p}(p)   \ ,
\end{eqnarray}
%-------------------------------------------------------------------------
where $\tau_{p}(p)$ is the polariton relaxation time in the normal phase given by
%-------------------------------------------------------------------------
\begin{eqnarray}
\label{tp} \frac{1}{\tau_{p}(p)} = \frac{2\pi}{\hbar} \int  \left|
V(\mathbf{p},\mathbf{p}^{\prime})
\right|^{2}X_{\mathbf{p}}^{2}X_{\mathbf{p}^{\prime}}^{2}
\delta(\varepsilon_{0}(p) - \varepsilon_{0}(p^{\prime})) \left(1 -
\cos(\hat{(\mathbf{p},\mathbf{p}^{\prime})}) \right) \frac{s
d^{2}p^{\prime}}{(2\pi\hbar)^{2}} \ .
\end{eqnarray}
%-------------------------------------------------------------------------
Hence, we obtain $\tau_{p}(p) =  \tau_{n}(p)/X_{\mathbf{p}}^{4}$,
where $\tau_{n}(p)$ is the exciton relaxation time in the normal
phase given by
%-------------------------------------------------------------------------
\begin{eqnarray}
\label{tpn}
\frac{1}{\tau_{n}(p)}  = \frac{2\pi}{\hbar} \int  \left| V(\mathbf{p},\mathbf{p}^{\prime}) \right|^{2} \delta(\varepsilon_{ex}(p) - \varepsilon_{ex}(p^{\prime})) \left(1 - \cos(\hat{(\mathbf{p},\mathbf{p}^{\prime})}) \right) \frac{s d^{2}p^{\prime}}{(2\pi\hbar)^{2}} \ ,
\end{eqnarray}
%-------------------------------------------------------------------------
where $\varepsilon_{ex}(p)$ is the energy spectrum of the excitons.
In the case of excitations with a small momentum, where
$\varepsilon_{0}(p) \ll \mu$, the dispersion law of the excitations
has an acoustic form: $\varepsilon_{1}(p) = c_{s}p$, and from
Eq.~(\ref{t12}) we have $\tau_{1}(p) = (p/(M_{p}c_{s}))\tau_{p}(p)$.
Therefore,
 in the presence of the superfluidity the relaxation time of polariton excitations  $\tau_{1}(p)$ can be obtained from the exciton normal phase relaxation
  time $\tau_{n}(p)$ as $\tau_{1}(p) = ( X_{\mathbf{p}}^{4} p/(M_{p}c_{s}))\tau_{n}(p)$. Without the superfluidity,
  at $\sigma (p_{1},p_{1}^{\prime}) = X_{\mathbf{p}_{1}}X_{\mathbf{p}_{1}^{\prime}}$ we have  $\tau_{1}(p) =  X_{\mathbf{p}}^{4} \tau_{n}(p)$.
  The exciton normal phase relaxation time $\tau_{n}(p)$ can be approximated by its average value $\bar{\tau}_{n} = \langle \tau_{n}(p)\rangle$, which
   can be obtained from the exciton mobility $\tilde{\mu}_{ex} = e \bar{\tau}_{n}/M$. The exciton mobility $\tilde{\mu}_{ex}$ is presented in Figs.~1 and~2 in Ref.~[\onlinecite{Basu}].

%-------------------------------------------------------------------------
%-------------------------------------------------------------------------

\section{The drag coefficients}
\label{drag_pe}

We introduce the drag coefficients $\lambda _{p}$ and $\lambda
_{ex}$ for electrons in the 2DEG dragged by the moving polaritons
and exitons, respectively. For the case when the electric field is
applied to the system of electrons in the QW we introduce the drag
coefficients $\gamma _{p}$ and $\gamma _{ex},$ respectively, for
polaritons and excitons dragged by the electron current. In the
two-layer system there are a current of electrons and
flow of polaritons or excitons. The flow of polaritons or excitons is $%
\mathbf{i}_{i}=n_{i}\mathbf{v}_{i}$, where $n_{i}$ and $v_{i}$ are
density and average velocity, and the index $i$ is defined as $i =
ex$ for excitons and $i = p$ for polaritons. The electron current $\mathbf{j}%
=-en_{el}\mathbf{v}_{el}$, where $n_{el}$ is\textbf{\ }the density
 and $\mathbf{v}_{el}$ is the average velocity of electrons in the electron layer.  These currents can be expressed in
terms of the density gradient $\mathbf{\nabla }n_{i}$ in the
polariton or exciton subsystem, drag coefficients $\lambda _{i}$ and
$\gamma _{i},$ and external electric field $\mathbf{E}$ applied to
the 2DEG by the following matrix expression:
%-------------------------------------------------------------------------
\begin{equation}
\left(
\begin{array}{c}
\mathbf{i}_{i} \\
\mathbf{j}%
\end{array}%
\right) =\left(
\begin{array}{cc}
-D_{i} & \gamma _{i} \\
\lambda _{i} & en_{el}D_{e}%
\end{array}%
\right) \cdot \left(
\begin{array}{c}
\mathbf{\nabla }n_{i} \\
\mathbf{E}%
\end{array}%
\right) \ , \label{matrix}
%-------------------------------------------------------------------------
\end{equation}%
where $D_{i}$ is the polariton or exciton diffusion coefficient and
$D_{e}$ is the mobility coefficient of the electrons. Only normal
component in the polariton subsystem is dragged by the electron
current, while the superfluid component is not dragged. Thus, the
appearance of the polariton superfluidity can be detected by the
electron-polariton drag effect.

%-------------------------------------------------------------------------
%-------------------------------------------------------------------------
%-------------------------------------------------------------------------
%-------------------------------------------------------------------------

\subsection{The drag of the polariton quasiparticles by the electron current.%
}

Let us find the drag coefficient $\gamma _{p}$ by following the
procedure applied in Ref.~[\onlinecite{Nikitkov2}] for the
derivation of the drag coefficient related to the drag of the
quasiparticles in the polariton subsystem by the electrons. We
obtain the drag coefficient $\gamma _{p}$ from the expansion of the
polariton flow $\mathbf{i}_{p}$ in the first order
with respect to $\mathbf{E}$. The expression for the polariton flow $\mathbf{%
i}_{p}$ is given by
%-------------------------------------------------------------------------
\begin{eqnarray}
\label{cur} \mathbf{i}_{p} = - \frac{1}{M_{p}} \int \mathbf{p}_{1}
n(\mathbf{p}_{1}) \frac{s d^{2}p_{1}}{(2\pi \hbar)^{2}}  \ ,
\end{eqnarray}
%-------------------------------------------------------------------------
where $\mathbf{p}_{1}$ is the polariton momentum, $s=4$ is the
degeneracy factor for polaritons, and $n(\mathbf{p}_{1})$ is the
distribution function of the quasiparticle excitations in the
polariton subsystem which can be found using kinetic equations. The
kinetic equations for distribution function of the quasiparticles
are represented in Appendix~\ref{ap.ke}.

Using the distribution function of the quasiparticles from Appendix~\ref%
{ap.ke}, we obtain the polariton flow $\mathbf{i}_{p}$ in the first
order with respect to external electric field $\mathbf{E}$, and find
the drag coefficient $\gamma_{p}$. As a result, we obtain for
$\gamma_{p}$:
%-------------------------------------------------------------------------
\begin{eqnarray}  \label{alpha}
\gamma_{p} &=& \frac{\pi}{2 \hbar} \frac{e}{M_{p}m_{e}k_{B}T} \int
\frac{s
d^{2}q}{(2\pi \hbar)^{2}} W_{eff}^{2}(q,D)  \nonumber \\
&\times& \int_{0}^{\infty} \frac{\tilde{\Phi}(\mathbf{q},\xi)\Psi(\mathbf{q}%
,\xi)}{\sinh^{2}(\xi/(2k_{B}T))} d\xi \ ,
\end{eqnarray}
%-------------------------------------------------------------------------
where
%-------------------------------------------------------------------------
\begin{eqnarray}  \label{Phi}
&& \tilde{\Phi}(\mathbf{q},\xi) = \int \sigma^{2} (p_{1},
|\mathbf{p}_{1} + \mathbf{q}|) [n_{0}(\varepsilon_{1}(\mathbf{p}_{1}
+ \mathbf{q}|)) -
n_{0}(\varepsilon_{1}(p_{1}))]  \nonumber \\
&&\times (\tau_{p}(|\mathbf{p}_{1} + \mathbf{q}|)(\mathbf{p}_{1} + \mathbf{q}%
) - \tau_{p}(p_{1})\mathbf{p}_{1}) \delta (\varepsilon_{1}(p_{1}) -
\varepsilon_{1}(|\mathbf{p}_{1} + \mathbf{q}|) + \xi) \frac{s d^{2}p_{1}}{%
(2\pi \hbar)^{2}}  \nonumber \\
&& \approx \frac{2s}{\pi\hbar^{2}c_{s}^{2}}\mathbf{q} \int \left.
\left( \sigma^{2} (p_{1}, |\mathbf{p}_{1} + \mathbf{q}|) \frac{
\partial n_{0}(\varepsilon_{1})}{\partial \varepsilon_{1}}\right)
\right|_{\varepsilon_{1} = \mu } (\varepsilon_{1}(|\mathbf{p}_{1} + \mathbf{q%
}|) - \varepsilon_{1}(p_{1})) \bar{\tau}_{n} \delta
(\varepsilon_{1}(p_{1}) - \varepsilon_{1}(|\mathbf{p}_{1} +
\mathbf{q}|) + \xi) \varepsilon_{1}
d\varepsilon_{1}  \nonumber \\
&& \approx -\frac{s\mu \bar{\tau}_{n} }{2\pi\hbar^{2}c_{s}^{2}k_{B}T}\frac{%
\exp[\mu/(k_{B}T)]}{(\exp[\mu/(k_{B}T)] -1)^{2}} \xi \mathbf{q} \ ,
\end{eqnarray}
%-------------------------------------------------------------------------
and
%-------------------------------------------------------------------------
\begin{eqnarray}  \label{Psi}
&& \Psi(\mathbf{q},\xi) = \int
[f_{0}(\varepsilon_{2}(|\mathbf{p}_{2} + \mathbf{q}|)) -
f_{0}(\varepsilon_{2}(p_{2}))] (\tau_{2}(|\mathbf{p}_{2} +
\mathbf{q}|)(\mathbf{p}_{2} + \mathbf{q}) -
\tau_{2}(p_{2})\mathbf{p}_{2})
\nonumber \\
&&\times \delta (\varepsilon_{2}(p_{2}) -
\varepsilon_{2}(|\mathbf{p}_{2} +
\mathbf{q}|) + \xi) \frac{2 \ d^{2}p_{2}}{(2\pi \hbar)^{2}}  \nonumber \\
&& \approx \frac{m_{e}}{\pi \hbar^{2}}\mathbf{q} \int \left. \left( \frac{%
\partial f_{0}(\varepsilon_{2})}{\partial \varepsilon_{2}} \right)
\right|_{\varepsilon_{2} = \varepsilon_{F}}
(\varepsilon_{2}(|\mathbf{p}_{2} + \mathbf{q}|) -
\varepsilon_{2}(p_{2})) \bar{\tau}_{2} \delta
(\varepsilon_{2}(p_{2}) - \varepsilon_{2}(|\mathbf{p}_{2} +
\mathbf{q}|) +
\xi) d\varepsilon_{2}  \nonumber \\
&& = - \frac{m_{e}\bar{\tau}_{2}}{4 \pi \hbar^{2}k_{B}T}\xi
\mathbf{q} \ .
\end{eqnarray}
%-------------------------------------------------------------------------
Assuming the system to be close to the equilibrium and considering
$\nabla \mu_{qp}(\mathbf{r})$ to be very small, we put
$\mu_{qp}(\mathbf{r}) = 0$ in Eqs.~(\ref{alpha})-~(\ref{Psi}).

\begin{figure}[t]
%  figure placement: here, top, bottom, or page
\centering
\includegraphics[width=3.5in]{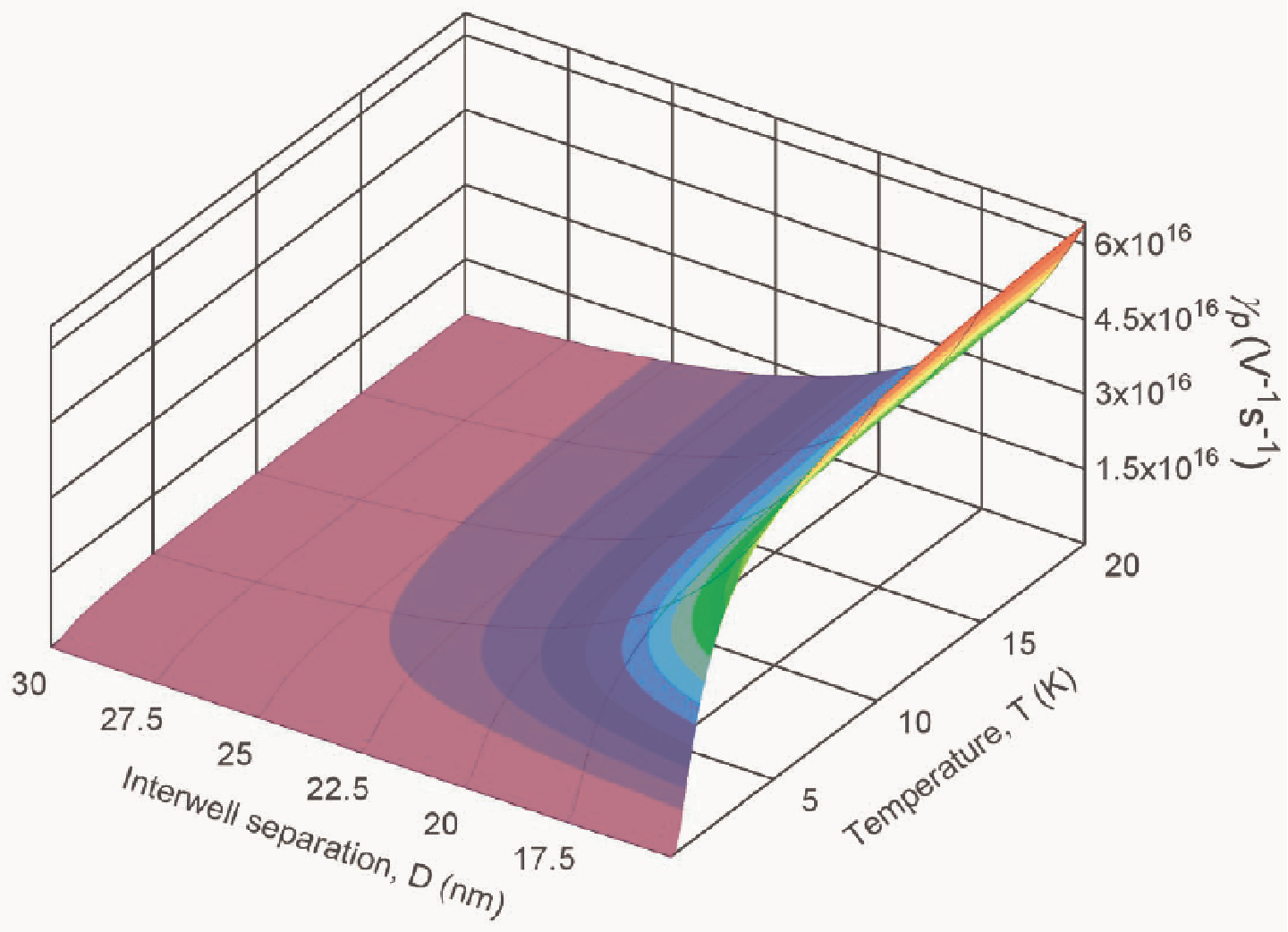}
\caption{(Color online) The drag coefficient $\protect\gamma_{p}$ in
$\mathrm{V^{-1}s^{-1}}$ in the system of superfluid microcavity
polaritons and electrons as a function of temperature $T$ in
$\mathrm{K}$ and interwell separation $D$ in $\mathrm{nm}$. The
polariton density $n_{p} = 10^{10} \ \mathrm{cm^{-2}}$. We used the
parameters for the GaAs/GaAsAl quantum wells:
$m_{e} = 0.07 m_{0}$, $m_{h} = 0.15 m_{0}$, $M = 0.24 m_{0}$, $\protect%
\epsilon = 13$.} \label{gam_lt1}
\end{figure}

\begin{figure}[t]
%  figure placement: here, top, bottom, or page
\centering
\includegraphics[width=3.5in]{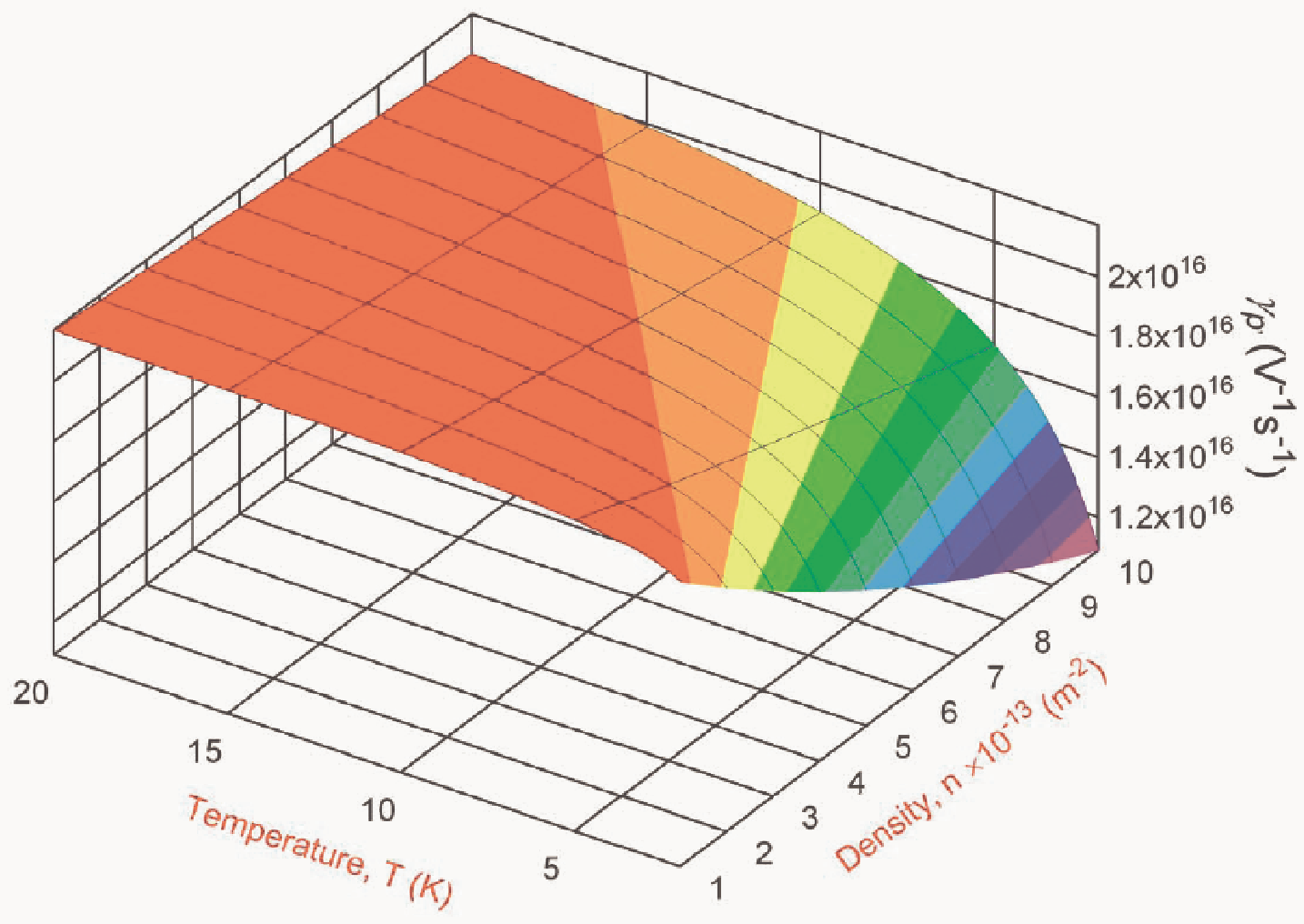}
\caption{(Color online) The drag coefficient $\protect\gamma_{pe}$
in $\mathrm{V^{-1}s^{-1}} $ in the system of superfluid microcavity
polaritons and electrons as a function of temperature $T$ in
$\mathrm{K}$ and polariton density $n_{p}$ in $\mathrm{m^{-2}}$. The
interwell separation $D = 20 \
\mathrm{nm}$. We used the parameters for the GaAs/GaAsAl quantum wells: $%
m_{e} = 0.07 m_{0}$, $m_{h} = 0.15 m_{0}$, $M = 0.24 m_{0}$, $\protect%
\epsilon = 13$.} \label{gam_lt2}
\end{figure}

Let us mention that the typical interwell distances used in the drag
experiment in Ref.~[\onlinecite{Gramila}] are $17.5 \ \mathrm{nm}$
and $22.5
\ \mathrm{nm}$. The interwell distances used in the experiment in Ref.~[%
\onlinecite{Lilly2}] are $20 \ \mathrm{nm}$ and $30 \ \mathrm{nm}$.
In our calculations we used the same interwell distances that used
in the drag experiments~\cite{Gramila,Lilly2}, namely $17.5 \
\mathrm{nm}$, $20 \
\mathrm{nm}$, $22.5 \ \mathrm{nm}$ and $30 \ \mathrm{nm}$. Figs.~\ref%
{gam_lt1} and~\ref{gam_lt2} present results of calculations for the
drag coefficient $\gamma_{p}$. The drag coefficient $\gamma_{p}$ as
a function of temperature $T$ and interwell separation $D$ is shown
in Fig.~\ref{gam_lt1}. The drag coefficient $\gamma_{p}$ as a
function of temperature $T$ and polariton density $n_{p}$ is
presented in Fig.~\ref{gam_lt2}. We can conclude that the drag
coefficient $\gamma_{p}$ exponentially decreases with the exciton
density $n$, exponentially increases with the temperature $T$ and
decreases with the interwell separation $D$.

Let us mention that in the GaAs quantum wells used in Ref.~[%
\onlinecite{Balili_prb}] the Rabi splitting is $13\ \mathrm{meV}$
($\sim
150\ \mathrm{K}$). Since the derivation $\gamma _{p}$ resulting in Eq.~(\ref%
{alpha}) implies low temperatures $k_{B}T\ll \mu $ and $k_{B}T\ll
\hbar
\Omega _{R}$, we can apply Eq.~(\ref{alpha}) for the temperatures below $%
\sim 20\ \mathrm{K}$.

Note that Eq.~(\ref{alpha}) was obtained by using the regular
Bogoliubov approximation for the weakly-interacting Bose gas with no
dissipation. If the exciton-polaritons have a finite life-time in
the cavity, the Bogoliubov dispersion is modified at small wave
vectors~\cite{Wouters}. This modification should affect the drag.
However, we consider the small relative deviation from the threshold
pumping intensity, when the pumping  maintains the exact balance of
amplification and losses. According to Eq.~(6) in
Ref.~[\onlinecite{Wouters}], at small relative deviation from the
threshold pumping intensity, the spectrum of collective excitations
in the system of exciton polaritons corresponds to the regular
Bogoliubov approximation. We assume that the leakage of the photons
from the microcavity is very small, and the system can be
considered in the quasi-equilibrium.

%%%%%%%%%%%%%%%%%%%%%%%%%%%%%%%%%%%%%%%%%%%%%%%%%%%%%%%%%%%%%%%%%%%%%%%%%%%%%%%%%%%%%%%%%%%%%%%%%%%%%%%%%%%%%%%%%%%%%%%%%%%%%%%%%%%%%%%%%%%%%%%%%%%
%%%%%%%%%%%%%%%%%%%%%%%%%%%%%%%%%%%%%%%%%%%%%%%%%%%%%%%%%%%%%%%%%%%%%%%%%%%%%%%%%%%%%%%%%%%%%%%%%%%%%%%%%%%%%%%%%%%%%%%%%%%%%%%%%%%%%%%%%%%%%%%%%%

%-------------------------------------------------------------------------
%-------------------------------------------------------------------------

\subsection{The drag of the electrons by the flow of polariton
quasiparticles.}

%\label{drag_ep}

Let us find the drag coefficient $\lambda_{p}$ related to the drag
of the electrons by the quasiparticles in the polariton subsystem.
We obtain the
drag coefficient $\lambda_{p}$ from the expansion of the electron current $%
\mathbf{j}$ in the first order with respect to $\mathbf{\nabla }
n_{qp}$, where $n_{qp}$ is the density of quasiparticles
contributing to the normal component of the polariton subsystem. The
expression for the electron current $\mathbf{j}$ is given by
%-------------------------------------------------------------------------
\begin{eqnarray}  \label{cur_e}
\mathbf{j} = - \frac{2 e}{m_{e}} \int \mathbf{p}_{2}
f(\mathbf{p}_{2}) \frac{ d^{2}p}{(2\pi \hbar)^{2}} \ ,
\end{eqnarray}
%-------------------------------------------------------------------------
where $\mathbf{p}_{2}$ is the electron momentum. and
$f(\mathbf{p}_{2})$ is the electron distribution function
represented by Eq.~(\ref{distel}).  We can find
$g_{2}(\mathbf{p}_{2})$ in Eq.~(\ref{distel}) by solving the kinetic
equations Eqs.~(\ref{kin1})-(\ref{kin2}).

Substituting Eq.~(\ref{born}) into Eq.~(\ref{i21}), we expand the
kinetic
equations~(\ref{kin1}) and~(\ref{kin2}) in the first order with respect to $%
\mathbf{\nabla}\mu_{qp}$, we find the functions $g_{1}$ and $g_{2}$.
Therefore, we can obtain the electron current $\mathbf{j}$ in the
first order with respect to $\mathbf{\nabla}n_{qp}$, and find the
drag coefficient $\lambda_{p}$. As a result, we obtain
$\lambda_{p}$:
%-------------------------------------------------------------------------
\begin{eqnarray}  \label{lambda}
\lambda_{p} &=& \frac{\pi}{2 \hbar} \frac{e}{M_{p}m_{e}k_{B}T} \left(\frac{%
\partial n_{qp}}{\partial \mu_{qp}}\right)^{-1} \int \frac{s d^{2}q}{(2\pi
\hbar)^{2}} W_{eff}^{2}(q,D)  \nonumber \\
&\times& \int_{0}^{\infty} \frac{\tilde{\Phi}(\mathbf{q},\xi)\Psi(\mathbf{q}%
,\xi)}{\sinh^{2}(\xi/(2k_{B}T))} d\xi \ ,
\end{eqnarray}
%-------------------------------------------------------------------------
where $\tilde{\Phi}(\mathbf{q},\xi)$ and $\Psi(\mathbf{q},\xi)$ are
given by Eqs.~(\ref{Phi}) and~(\ref{Psi}), correspondingly.

Applying  for the density of the quasiparticles $n_{qp}$
%-------------------------------------------------------------------------
\begin{eqnarray}  \label{npq}
n_{qp} = \int \frac{1}{\exp([\varepsilon_{1}(\mathbf{p}_{1}) -\mu_{qp}(%
\mathbf{r})] /(k_{B}T)) - 1}\frac{s d^{2}p_{1}}{(2\pi \hbar)^{2}} \
,
\end{eqnarray}
%-------------------------------------------------------------------------
we obtain $(\partial n_{qp})/(\partial \mu_{qp})$:
%-------------------------------------------------------------------------
\begin{eqnarray}  \label{der1}
&& \left.\left(\frac{\partial n_{qp}}{\partial \mu_{qp}}\right)\right|_{%
\mu_{qp}=0} = \int \left. \frac{\partial }{\partial \mu_{qp}}%
\right|_{\mu_{qp}=0} \left(
\frac{1}{\exp([\varepsilon_{1}(\mathbf{p}_{1})
-\mu_{qp}(\mathbf{r})] /(k_{B}T)) - 1}\right)\frac{s
d^{2}p_{1}}{(2\pi
\hbar)^{2}}  \nonumber \\
&& = \frac{1}{k_{B}T}\int\frac{\exp[\varepsilon_{1}(\mathbf{p}_{1})/(k_{B}T)]%
}{\left(\exp[\varepsilon_{1}(\mathbf{p}_{1})/(k_{B}T)] - 1\right)^{2}}\frac{%
s d^{2}p_{1}}{(2\pi \hbar)^{2}} = \frac{s k_{B}T}{2\pi\hbar^{2}c_{s}^{2}}%
\int_{0}^{\infty} \frac{e^{x}x dx}{(e^{x}-1)^{2}} \ ,
\end{eqnarray}
%-------------------------------------------------------------------------
where we use $\varepsilon_{1}(\mathbf{p}_{1}) = c_{s}p_{1}$ and $x =
c_{s}p_{1}/k_{B}T$ for the small momenta. The integral in the r.h.s. of Eq.~(%
\ref{der1}) diverges at $p\rightarrow 0$. Therefore, we have
$(\partial n_{qp})/(\partial \mu_{qp}) \rightarrow \infty$, which
results in $\lambda _{p} = 0$ according to Eq.~(\ref{lambda}). This
result comes from the assumption that for the very dilute Bose gas
of polaritons we took into account only the sound region of the
collective excitation spectrum at small momenta, and neglect almost
not occupied regions with the quadratic spectrum at large momenta
and crossover dependence of the spectrum at the
intermediate momenta. Our approximation results in suppressed $\lambda _{p}$%
. Therefore, in the presence of the superfluidity of polaritons, the
polaritons moving due to their density gradient almost do not drag
electrons and there is suppressed electron current induced by the
polaritons. Hence, the suppression of the dragged electric current
in the electron QW can indicate the superfluidity of the polaritons.

%-------------------------------------------------------------------------
%-------------------------------------------------------------------------

\section{The drag effects in the exciton-electron system at high temperatures%
}

\label{drag_exe}

For high temperature, $k_{B}T\gtrsim \hbar \Omega _{R}$, the majority of polaritons occupy the
upper polariton branch, where the upper polariton mass is very close to the mass of exciton
$M_{ex}$. So at high temperature the polaritons are replaced by the gas of excitons~\cite{Lit}.

Without the superfluidity in the definitions of the drag
coefficients presented by Eq.~(\ref{matrix}) it should be
substituted the exciton density $n_{ex}$ instead of the
quasiparticle density $n_{qp}$, exciton mass $M_{ex}$
instead of polariton mass $M_{p}$, and the chemical potential of excitons $%
\mu_{ex}$ instead of the chemical potential of the quasiparticles
$\mu_{qp}$ The drag coefficients $\gamma_{ex}$ for the exciton
system without the superfluidity can be obtained from
Eqs.~(\ref{alpha})-~(\ref{Psi}) by substituting $\sigma
(p_{1},p_{1}^{\prime}) = 1$, $\varepsilon_{1}(p_{1}) =
p_{1}^{2}/(2M_{ex})$, and $n_{0}(\mathbf{p}_{1})) =
\left(\exp([\varepsilon_{1}(\mathbf{p}_{1})
-\mu_{ex}^{(0)}(\mathbf{r})] /(k_{B}T)]) - 1\right)^{-1}$ is the
Bose-Einstein distribution function of the excitons at the
equilibrium, where $\mu_{ex}^{(0)}$ is the chemical potential of the
excitons in the equilibrium determined by the polariton density
$n_{ex}$:
%-------------------------------------------------------------------------
\begin{eqnarray}  \label{npm}
n_{ex} = \int \frac{1}{\exp([\varepsilon_{1}(\mathbf{p}_{1})
-\mu_{ex}^{(0)}] /(k_{B}T)) - 1}\frac{s d^{2}p_{1}}{(2\pi
\hbar)^{2}} \ .
\end{eqnarray}
%-------------------------------------------------------------------------
From Eq.~(\ref{npm}) we obtain the $\mu_{ex}^{(0)}$:
%-------------------------------------------------------------------------
\begin{eqnarray}  \label{mup}
\mu_{ex}^{(0)} = k_{B}T \log \left[1 - \exp\left[-\frac{2\pi
\hbar^{2}n}{s M_{ex} k_{B}T}\right] \right] \ .
\end{eqnarray}
%-------------------------------------------------------------------------
We get from Eq.~(\ref{npm}):
%-------------------------------------------------------------------------
\begin{eqnarray}  \label{npm2}
n_{ex} = - \frac{s M_{ex} k_{B}T}{2\pi\hbar^{2}} \left(1 - \exp\left[\frac{%
\mu_{ex}(\mathbf{r})}{k_{B}T} \right] \right) \ .
\end{eqnarray}
%-------------------------------------------------------------------------
From Eq.~(\ref{npm2}) we find:
%-------------------------------------------------------------------------
\begin{eqnarray}  \label{npm3}
\left.\left(\frac{\partial n_{ex}}{\partial \mu_{ex}}\right)\right|_{%
\mu_{ex}=\mu_{ex}^{(0)}} = \frac{s M_{ex}}{2\pi\hbar^{2}} \left[1 - \exp%
\left[-\frac{2\pi \hbar^{2}n}{s M_{ex} k_{B}T}\right] \right] \ .
\end{eqnarray}
%-------------------------------------------------------------------------

The coefficient $\lambda_{ex}$ for the high temperature range is
given by
%-------------------------------------------------------------------------
\begin{eqnarray}  \label{rel111}
\lambda_{ex} &=& \left(\left.\frac{\partial n_{ex}}{\partial \mu_{ex}}%
\right|_{\mu_{ex} = \mu_{ex}^{(0)}}\right)^{-1}\frac{\pi}{2 \hbar} \frac{e}{%
M_{ex}m_{e}k_{B}T} \int \frac{s d^{2}q}{(2\pi \hbar)^{2}}
W_{eff}^{2}(q,D)
\nonumber \\
&\times& \int_{0}^{\infty} \frac{\tilde{\Phi}(\mathbf{q},\xi)\Psi(\mathbf{q}%
,\xi)}{\sinh^{2}(\xi/(2k_{B}T))} d\xi \ ,
\end{eqnarray}
%-------------------------------------------------------------------------
where $\tilde{\Phi}(\mathbf{q},\xi)$ and $\Psi(\mathbf{q},\xi)$ are
defined
by the the substituting ex excitonic parameters described above to Eqs.~(\ref%
{Phi}) and~(\ref{Psi}), correspondingly.

\begin{figure}[t]
%  figure placement: here, top, bottom, or page
\centering
\includegraphics[width=3.5in]{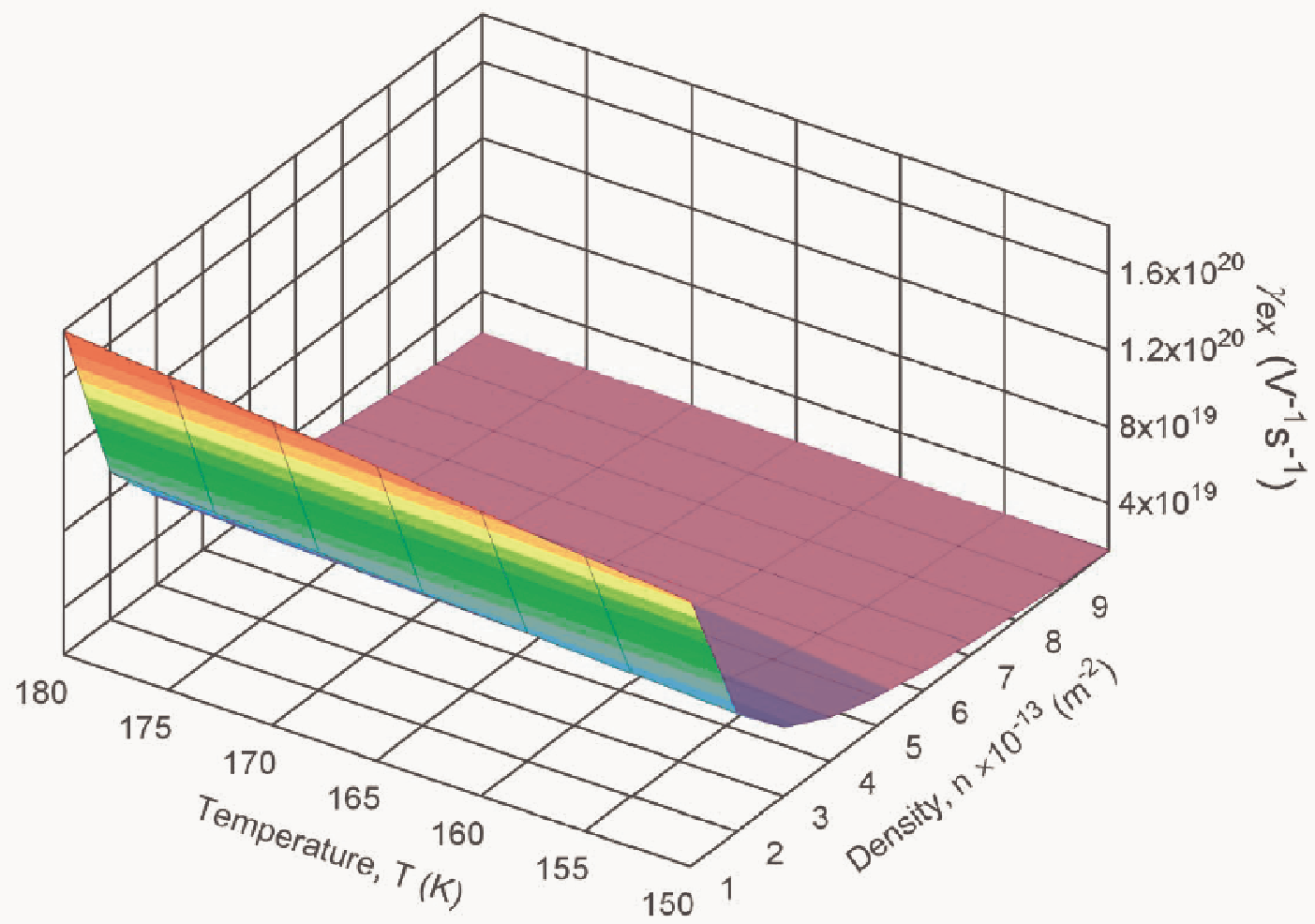}
\caption{(Color online) The drag coefficient $\protect\gamma_{exe}$ in $\mathrm{V^{-1}s^{-1}%
}$ in the system of spatially separated excitons and electrons
without superfluidity as a function of temperature $T$ in
$\mathrm{K}$ and exciton density $n_{ex}$ in $\mathrm{m^{-2}}$. The
interwell separation is $D = 20 \
\mathrm{nm}$. We used the parameters for the GaAs/GaAsAl quantum wells: $%
m_{e} = 0.07 m_{0}$, $m_{h} = 0.15 m_{0}$, $M = 0.24 m_{0}$, $\protect%
\epsilon = 13$.} \label{gam ht1}
\end{figure}

\begin{figure}[t]
%  figure placement: here, top, bottom, or page
\centering
\includegraphics[width=3.5in]{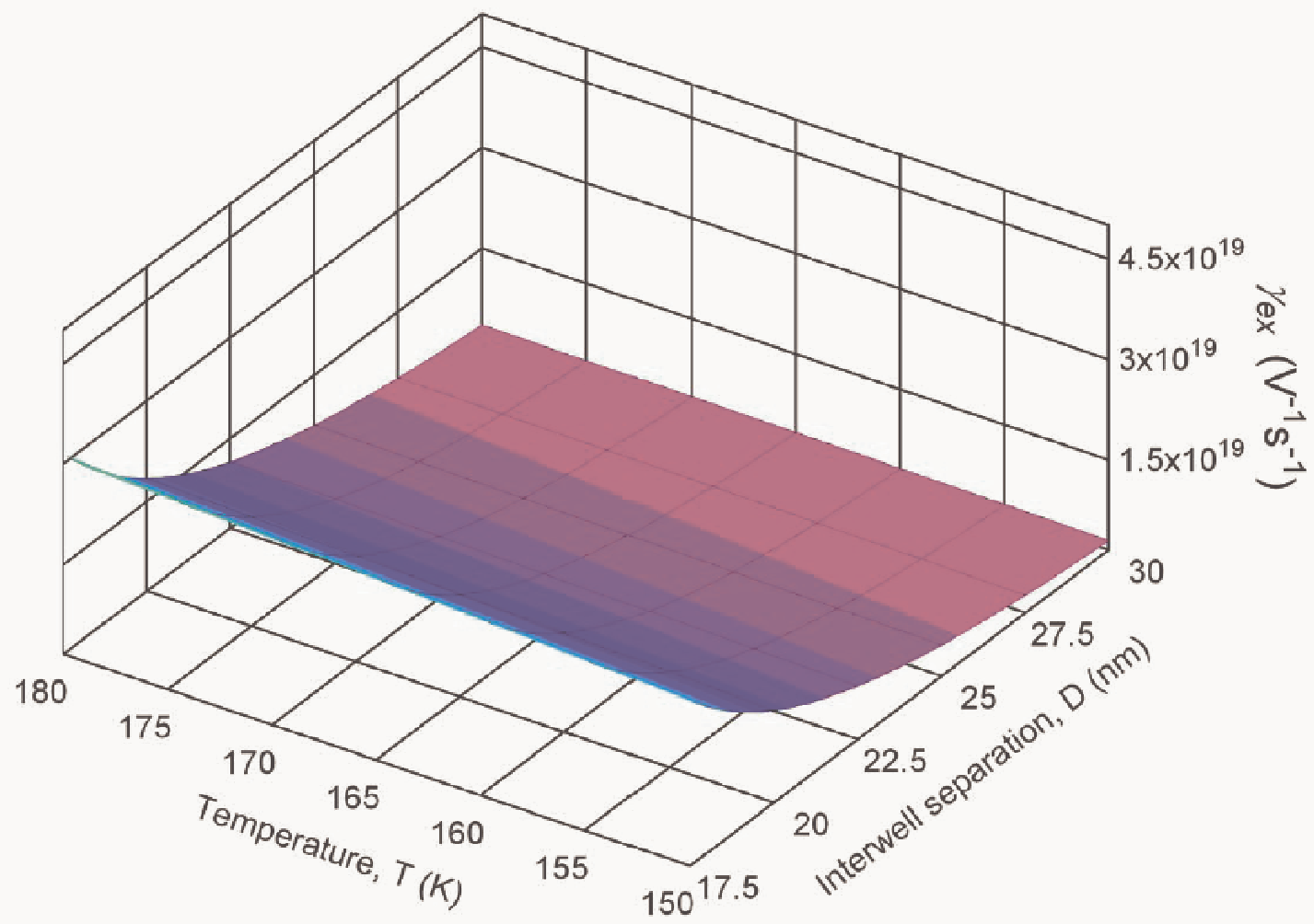}
\caption{(Color online) The drag coefficient $\protect\gamma_{exe}$ in $\mathrm{V^{-1}s^{-1}%
}$ in the system of spatially separated excitons and electrons
without superfluidity as a function of temperature $T$ in
$\mathrm{K}$ and interwell separation $D$ in $\mathrm{nm}$. The
exciton density $n_{ex} = 10^{10} \ \mathrm{cm^{-2}}$. We used the
parameters for the GaAs/GaAsAl quantum wells:
$m_{e} = 0.07 m_{0}$, $m_{h} = 0.15 m_{0}$, $M = 0.24 m_{0}$, $\protect%
\epsilon = 13$.} \label{gam ht2}
\end{figure}

\begin{figure}[t]
%  figure placement: here, top, bottom, or page
\centering
\includegraphics[width=3.5in]{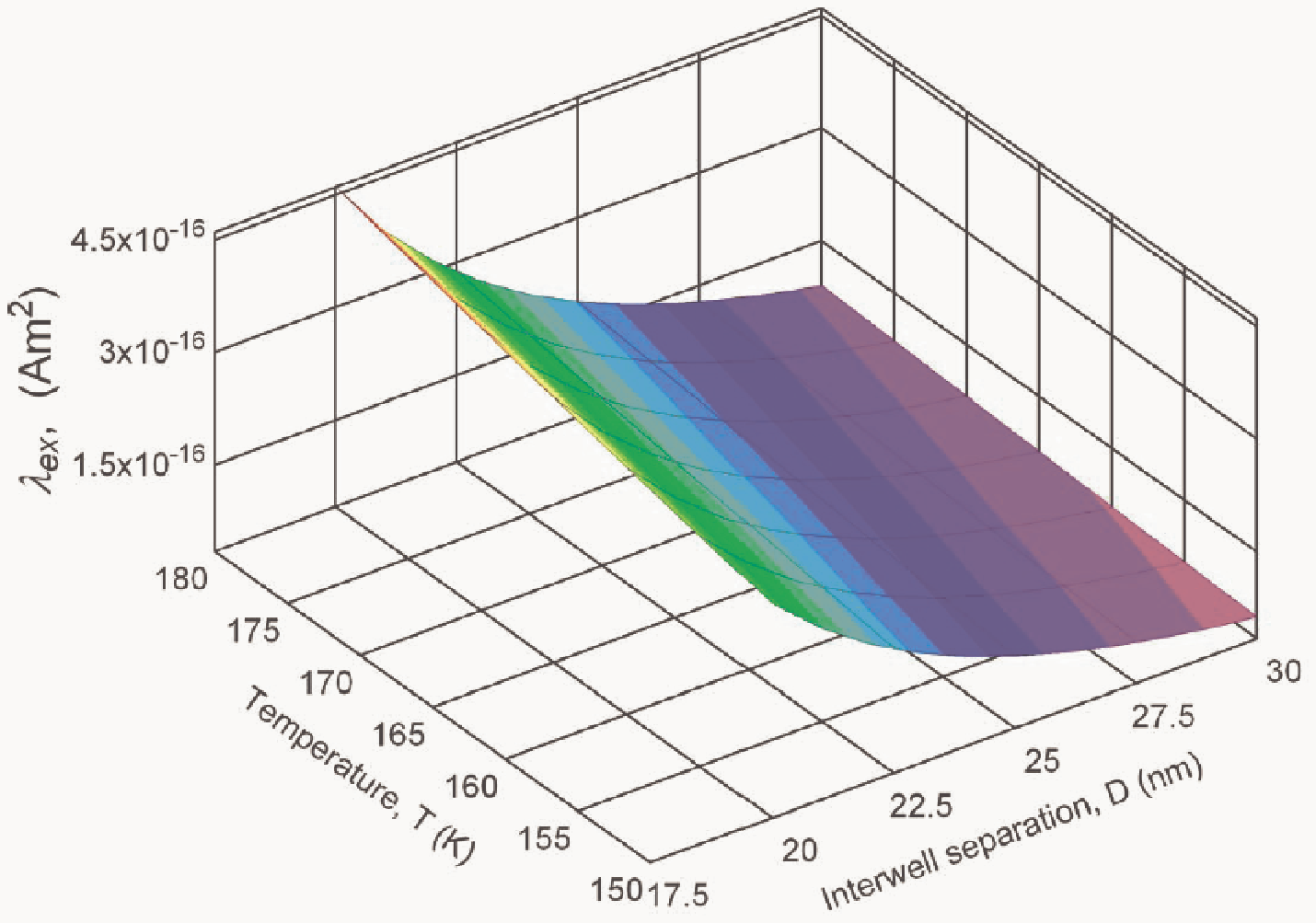}
\caption{(Color online) The drag coefficient $\protect\lambda
_{eex}$ in $\mathrm{A \ m^{2}} $ in the system of spatially
separated excitons and electrons without superfluidity as a function
of temperature $T$ in $\mathrm{K}$ and interwell separation $D$ in
$\mathrm{nm}$. The exciton density $n_{ex} = 10^{10} \
\mathrm{cm^{-2}}$. We used the parameters for the GaAs/GaAsAl
quantum wells:
$m_{e} = 0.07 m_{0}$, $m_{h} = 0.15 m_{0}$, $M = 0.24 m_{0}$, $\protect%
\epsilon = 13$.} \label{lam ht1}
\end{figure}

The results of the calculations of the drag coefficients at high
temperature are presented in Figs.~~\ref{gam ht1},~\ref{gam ht2}
and~~\ref{lam ht1}. The drag coefficient $\gamma_{ex}$ as a function
of temperature and exciton
density is shown in Fig.~\ref{gam ht1}, while the drag coefficient $%
\gamma_{ex}$ as a function of temperature and interwell separation
is shown in Fig.~\ref{gam ht2}. The drag coefficient $\lambda_{ex}$
at the high temperature range for electrons dragged by moving
excitons as a function of temperature and interwell separation is
presented in Fig.~\ref{lam ht1}. Based on the results of
calculations we can conclude that the drag coefficients
$\gamma_{ex}$ and $\lambda _{ex}$ decrease with the exciton density,
increase with the temperature and decrease with the interwell
separation.

%-------------------------------------------------------------------------
%-------------------------------------------------------------------------

\section{Proposed experiments}

\label{experiment}

We propose the following experiments relevant to electron-polariton and electron-exciton drag
effect. Let the screen with two diaphragms covers the quantum well embedded into a semiconductor
microcavity. The laser pumping through one diaphragm generates excitons forming the polaritons by
coupling to the cavity photons. At low temperature regime the absence of the electron current
will indicate the superfluidity phase of the polaritons, while at high temperature regime the
existence of the electron current will indicate the drag of electrons by the moving excitons.
Therefore, the drag coefficient $\lambda _{ex}$ allows to estimate the dragged electron current.
When the electron current is induced by the external electric field applied to the electron QW,
the photoluminescence spectrum can be measured in the other diaphragm. At low temperature, the
difference between the photoluminescence spectrum of polaritons decay with the electric field
applied to the electrons and without the electric field will indicates that the polaritons moved
to the other place of the QW due to the drag by the electron current. The photoluminescence
spectrum in the other diaphragm without the electric field is caused only by the diffusion of the
polaritons. Only the normal component of the polariton subsystem will move to the other
diaphragm, while the superfluid component is not affected by the electrons. It seems like
electrons move the photons that coupled with excitons. At high temperature regime the
photoluminescence spectrum of the electron-hole recombination indicates that the excitons moved
to the other place of the QW due to the drag by the electron current.

\begin{figure}[t]
%  figure placement: here, top, bottom, or page
\centering
\includegraphics[width=3.5in]{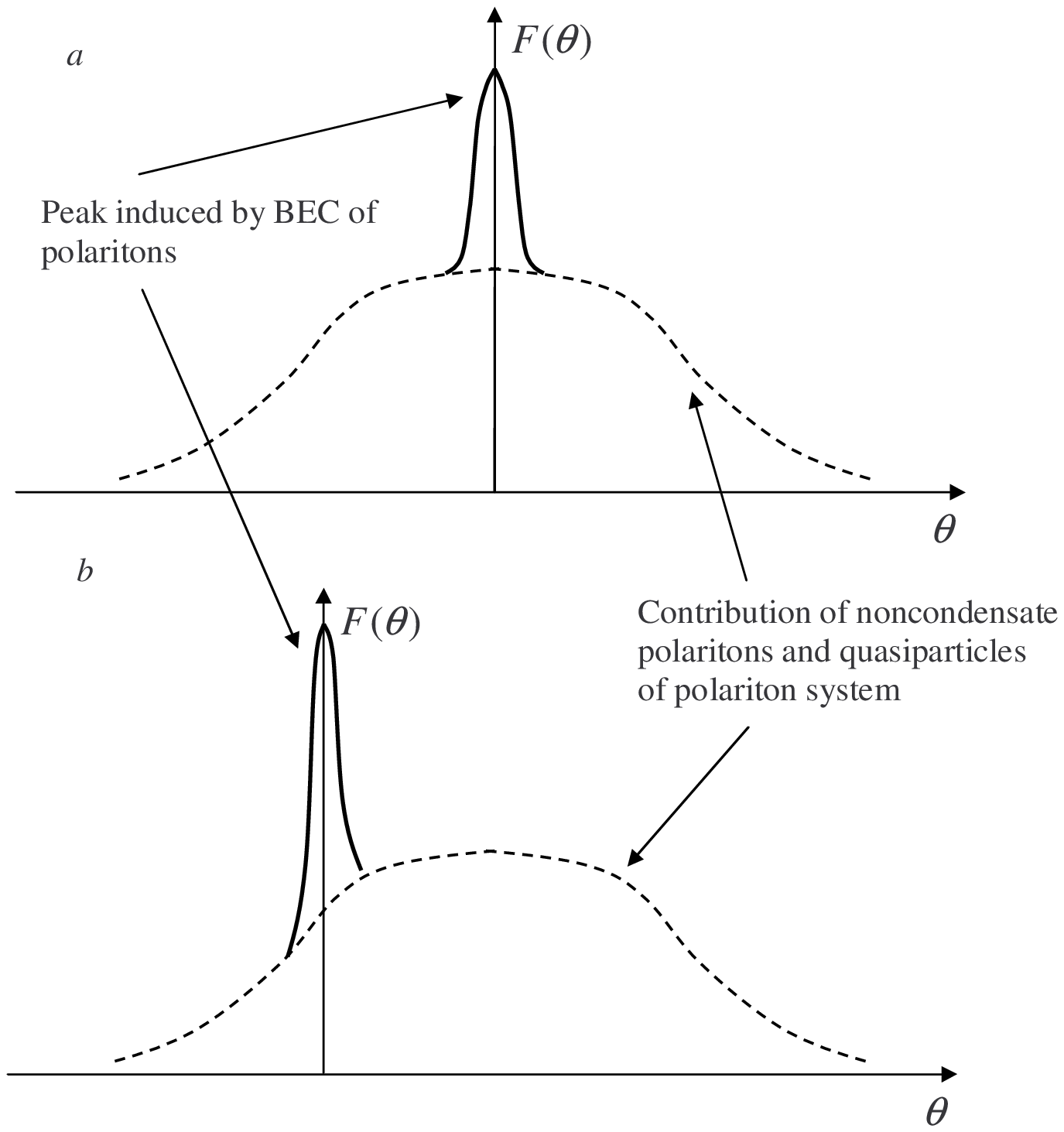}
\caption{Proposed experiment: the manifestation of polariton drag
effect through change of the angular distribution of the photons
escaping the optical microcavity. a. The angular distribution of the
photons escaping the optical microcavity without drag. b. The
angular distribution of the photons escaping the optical microcavity
in the presence of the drag effect. Redistribution of the
contribution of noncondensate polaritons.} \label{exper}
\end{figure}

The other suggested experiment is based on the observation of the
angular distribution of the photons escaping the optical
microcavity. At low temperature regime we propose to create the
uniform distribution of
polaritons by the laser pumping within the microcavity. Therefore, $\mathbf{%
\nabla }n_{p}=0$ and there is no polariton flow. In the absence of
polariton flow the average angle of the photons escaping the optical
microcavity and the perpendicular to the microcavity is
$\bar{\alpha}=0$, because the angular distribution is
symmetrical~\cite{science}. Let us induce the electron current by
applying the electric field $\mathbf{E}$ and analyze the photon
angular distribution in the presence of the nonzero current of
polaritons along the quantum well parallel to the cavity dragged by
the electron current due to the drag effect. If $\ \mathbf{\nabla
}n_{p}=0$ the polariton flow is $\mathbf{i}_{p}=\gamma
_{p}\mathbf{E}$, and according to
the definition of polariton flow we have $\mathbf{v}_{p}=\gamma _{p}\mathbf{E%
}/n_{p}$. Therefore, we can obtain the average component of the
polariton
momentum in the direction parallel to the Bragg mirrors of the microcavity: $%
\overline{p_{||}}=M_{p}v_{p}=M_{p}\gamma _{p}E/n_{p}$. Since the
perpendicular to the Bragg mirrors component of the polariton
momentum is given by~$p_{\bot }=\hbar \pi /L_{C}$ \cite{Snoke_text},
we obtain for the average tangent of the angle between the path of
the escaping photon and the perpendicular to the microcavity:
%-------------------------------------------------------------------------
\begin{eqnarray}
\label{tan} \overline{\tan \ \alpha
}=\frac{\overline{p_{||}}}{p_{\bot }}=\frac{\gamma _{p}
M_{p}L_{C}E}{\hbar \pi n_{p}}\ .
\end{eqnarray}
%-------------------------------------------------------------------------
Note that only normal component of the polariton subsystem will
contribute to the drag coefficient $\gamma _{p}$ and therefore to
$\overline{\tan \
\alpha }$. There will be two peaks of the escaping photons: one at $%
\overline{\tan \ \alpha }\neq 0$ corresponds to the moving (dragged)
normal component, and the other one at $\overline{\tan \ \alpha }=0$
corresponds to the superfluid component. Note that the analysis of
the angular distributions of the photons escaping the optical
microcavity has been used in the experiments~\cite{Amo1,Amo2}. The
manifestation of polariton drag effect through the change of the
angular distribution of the photons escaping the optical microcavity
is shown in Fig.~\ref{exper}. Only quasiparticles in polariton
system are dragged by electrons.

Let us make estimations of the parameters for the drag effects. At
the temperature $T = 4 \ \mathrm{K}$ the
experiment~\cite{science,Balili_prb} shows that the polariton
lifetime $\tau = 10 \ \mathrm{ps}$, and the polariton diffusion path
is $l = 20 \ \mathrm{\mu m}$. The corresponding
average polariton velocity is $v_{p} = l/\tau = 2 \times 10^{6} \ \mathrm{m/s%
}$. Since $E= n_{p}v_{p}/\gamma_{p}$, we can estimate the electric
field $E$ corresponding to such drag effect. For $n_{p} = 10^{10} \
\mathrm{cm^{-2}}$ and $T = 4 \ \mathrm{K}$ for the interwell
separation $D = 17.5 \ \mathrm{nm} $ $\gamma_{p} = 2.64 \times
10^{16} \ \mathrm{V^{-1}s^{-1}}$, the corresponding electric field
is $E = 3.8 \times 10^{3} \ \mathrm{V/m}$ which corresponds to the
applied voltage $V = 3.8 \times 10^{-3} \ \mathrm{V} $ at the size
of the system $d = 1 \ \mathrm{\mu m}$. Using $M_{p} = 7 \times
10^{-5} \times m_{e}$, the length of the microcavity $L_{C} = 2 \ \mathrm{%
\mu m}$~\cite{science,Balili_prb}, and the estimated $\gamma_{p}$
and $E$
in~(\ref{tan}), we obtain  for the average tangent of the angle $\alpha$: $%
\overline{\tan \ \alpha} = 0.385$ and $\tan^{-1}\left(\overline{\tan \ \alpha%
} \right) = 21^{0}$.

%-------------------------------------------------------------------------
%-------------------------------------------------------------------------

\section{Discussion and conclusions}

\label{disc}

Let us mention that we calculated the drag coefficients as the
linear responses of the equilibrium system, and, therefore, our
formulas for the drag coefficients are expressed in terms of the
parameters of the system at the equilibrium.

The other approximation used in our approach is that we considered
the polariton-electron drag only due to exciton-electron
interaction, neglecting the contribution coming from the cavity
photon-electron interaction. There are several processes caused by
the photon-electron interaction. The magnitude of the direct cavity
photon-electron static scattering processes are the second order
with respect to the fine-structure constant, and, therefore, their
contributions are much smaller than the contribution caused by the
exciton-electron interaction. The frequency of the plasmon
excitations in 2DEG are proportional to $k^{1/2}$, where $k$ is the
wave
vector, and the characteristic $k$ is by the order of magnitude of $%
2m_{e}e^{2}/(\hbar^{2}\epsilon)$ due to the screening~\cite{AFS}.
Therefore, the characteristic frequencies corresponding to the
plasmon excitations in 2DEG are in THz range, while the
characteristic frequencies of the cavity
photons are in the optical range by the magnitude of $10^{15} \ \mathrm{Hz}$%
. Thus the plasmon excitations in 2DEG are not in the resonance with
the cavity photons, while we consider polaritons formed by the
excitons and cavity photons in the resonance. The electron
intersubband transitions corresponding to the transitions between
the different levels of the quantization in the QW are also not in
the resonance with the cavity photons, because the characteristic
width of the QW $d$ managing the frequency of these intersubband
transitions is much smaller than the size of microcavity $L_{C}$
managing the frequency of the cavity photons. The frequencies of the
electron interband transitions in the semiconductor corresponding to
the range in the continuous electron spectrum, where there is no
discrete levels, also cannot be in the resonance with the
microcavity photons. The screening of the microcavity photon
spectrum by the 2DEG causes shift in the spectrum of microcavity
photons, and therefore, the frequencies corresponding to the
exciton-photon resonance related to the formation of the polariton
are shifted. However, the formalism and procedure for the
calculation of the polariton-electron drag coefficients will be the
same as presented in this Paper. The rigorous analysis of the
electron-microcavity photon drag effects is the very interesting
direction for the future research.

We can conclude that for systems of spatially separated interacting quasiparticles is the
possibility of controlling the motion of the quasiparticles of one subsystem by altering the
parameters of state of the quasiparticles in the other subsystem, for example, controlling the
flow of polaritons or exciton using a current of electrons. At low temperatures the electron
current dragged by the polariton flow is strongly suppressed and hence, the absence of the
electron current indicates the superfluidity of polaritons. However, the polariton flow can be
dragged by the electrons, and, therefore, there is a transport of photons along the microcavity,
which decreases wth rise of the superfluid component and can be observed through the change in
angular distribution of photons discussed above. At high temperatures, from one hand, the
existence of the electric current in the electron QW indicates the exciton flow in the other QW,
and from the other hand, the electron current in one QW induces the exciton flow in the other QW
via the drag of excitons by the electrons. The obtained drag coefficients allow calculate the
corresponding currents. According to our calculations, the low temperature drag coefficient
$\gamma _{p}$ and the high temperature drag coefficients $\gamma _{ex}$ and $\lambda _{ex}$
decrease with the exciton density, increase with the temperature and decrease with the interwell
separation. The suggested experiments allow to observe the analyzed drag effects. We suggested
the experiment for the observation of the distributions of the angles $\alpha $ between the path
of the photons escaping the microcavity and the perpendicular to the Bragg mirrors. The average
tangent of the angle $\overline{\tan \ \alpha }$ between the path of the escaping photon and the
perpendicular to the micocavity is proportional to the drag coefficient $\gamma _{p}$ and the
electric field $\mathbf{E}$ applied to the electrons.

%-----------------------------------------------------------------------------------------------------------------------
%-----------------------------------------------------------------------------------------------------------------------
\acknowledgments

O.~L.~B., R.~Ya.~K. were supported by PSC CUNY grant 63443-00 41
 and grant 62136-00 40; Yu.~E.~L. and A.~A.~K. were
supported by RFBR and RAS programs.
%-----------------------------------------------------------------------------------------------------------------------
%-----------------------------------------------------------------------------------------------------------------------

\appendix

%====================================================================================
%==================================

\section{The quasiparticle operators representation for the Hamiltonian of
the exciton-electron interaction}

\label{ap.qp}

Let us express the exciton operators in terms of polariton
operators. The exciton and photon operators are defined as
\cite{Ciuti}
%-------------------------------------------------------------------------
\begin{eqnarray}
\label{bog_tr} \hat{a}_{\mathbf{p}} = X_{p}\hat{l}_{\mathbf{p}} -
C_{p}\hat{w}_{\mathbf{p}} \ , \hspace{3cm}
 \hat{d}_{\mathbf{p}} = C_{p}\hat{l}_{\mathbf{p}} +
X_{p}\hat{w}_{\mathbf{p}} \ ,
\end{eqnarray}
%-------------------------------------------------------------------------
where $\hat{l}_{\mathbf{p}}$ and $\hat{w}_{\mathbf{p}}$ are lower
and upper polariton Bose operators, respectively, $X_{p}$ and
$C_{p}$ are
%-------------------------------------------------------------------------
\begin{eqnarray}
\label{bog} X_{p} = \left(1 +
\left(\frac{\hbar\Omega_{R}}{\varepsilon_{LP}(p) -
 \varepsilon _{ph}(p)}\right)\right)^{-1/2} \ , \hspace{3cm}
C_{P} = - \left(1 + \left(\frac{\varepsilon_{LP}(p) -
 \varepsilon _{ph}(p)}{\hbar\Omega_{R}} \right)\right)^{-1/2} \ ,
\end{eqnarray}
%-------------------------------------------------------------------------
and the energy spectra of the low/upper polaritons are
%-------------------------------------------------------------------------
\begin{eqnarray}
\varepsilon _{LP/UP}(p) &=&\frac{\varepsilon _{ph}(p)+\varepsilon _{ex}(p)}{2%
}  \nonumber  \label{eps0} \\
&\mp &\frac{1}{2}\sqrt{(\varepsilon _{ph}(p)-\varepsilon
_{ex}(p))^{2}+4|\hbar \Omega _{R}|^{2}}\ .
\end{eqnarray}%
%-------------------------------------------------------------------------
Eq.~(\ref{eps0}) implies a splitting between the upper and lower
states of polaritons at $p=0$ of $2\hbar \Omega _{R}$, known as the
Rabi splitting. Let us also mention that $|X_{p}|^{2}$ and
$|C_{p}|^{2}=1-|X_{p}|^{2}$ represent the exciton and cavity photon
fractions in the lower polariton.

In Eq.~ (\ref{eps0}), $\varepsilon_{ex}(p)$ is the energy dispersion
of a single exciton in a quantum well given by
%-------------------------------------------------------------------------
\begin{eqnarray}  \label{sp_ex}
\varepsilon_{ex}(p) = E_{band} - E_{binding} +
\tilde{\varepsilon}_{0}(p) \ ,
\end{eqnarray}
%-------------------------------------------------------------------------
where $E_{band}$ is the band gap energy, $E_{binding} =
\mu_{0}e^{4}/(\hbar^{2}\epsilon)$ is the binding energy of a 2D
exciton, and $\tilde{\varepsilon}_{0}(p) = p^2/(2M_{ex})$, where
$M_{ex} = m_{e} + m_{h}$ is the mass of an exciton. The cavity
photon spectrum is given by
%-------------------------------------------------------------------------
\begin{eqnarray}  \label{sp_phot}
\varepsilon _{ph}(p) = (c/\tilde{n})\sqrt{p^{2} +
\hbar^{2}\pi^{2}L_{C}^{-2}} \ .
\end{eqnarray}
%-------------------------------------------------------------------------
In Eq.~(\ref{sp_phot}), $c$ is the speed of light, $L_{C}$ is the
length of the cavity, $\tilde{n} = \sqrt{\epsilon}$ is the effective
refractive index. We assume the length of microcavity has the
following form
%-------------------------------------------------------------------------
\begin{eqnarray}  \label{lc}
L_{C} = \frac{\hbar\pi c}{\tilde{n} \left(E_{band} - \mathcal{E}%
_{binding}\right)} \ .
\end{eqnarray}
%-------------------------------------------------------------------------
So the photonic and excitonic branches start at the resonance at $p
= 0$.
This means that $\varepsilon_{ex}(p) = \varepsilon_{ph}(p)$ at $p = 0$ if $%
L_{C}$ satisfies to Eq.~(\ref{lc}).

At small momenta $\alpha \equiv 1/2 (M_{ex}^{-1} + (c/\tilde{n}%
)L_{C}/\hbar\pi)p^{2}/|\hbar \Omega_{R}| \ll 1$, the single-particle
lower
polariton spectrum obtained by substitution of Eq.~(\ref{sp_ex}) into Eq.~(%
\ref{eps0}), in linear order with respect to the small parameters
$\alpha$, is
%-------------------------------------------------------------------------
\begin{eqnarray}  \label{eps00}
\varepsilon_{0}(p) \approx \frac{c}{\tilde{n}} \hbar \pi L_{C}^{-1}
- |\hbar \Omega_{R}| + \frac{\gamma}{4} r ^{2} + \frac{1}{4}
\left(M_{ex}^{-1} + \frac{c L_{C}}{\tilde{n}\hbar\pi}\right)p^{2} =
\frac{p^{2}}{2M_{p}} \ ,
\end{eqnarray}
%-------------------------------------------------------------------------
where $M_{p}$ is the effective mass of polariton given by
%-------------------------------------------------------------------------
\begin{eqnarray}  \label{Meff}
M_{p} = 2 \left(M_{ex}^{-1} + \frac{c
L_{C}}{\tilde{n}\hbar\pi}\right)^{-1} \ .
\end{eqnarray}
%-------------------------------------------------------------------------

Substituting Eq.~(\ref{bog_tr}) into Eq.~(\ref{Ham_exc}), and taking
into account only the lower polaritons corresponding to the lower
energy, we obtain the Hamiltonian $\hat{H}_{ex-el}$ expressed in
terms of the lower polariton operators:
%-------------------------------------------------------------------------
\begin{eqnarray}  \label{Ham_pol}
\hat{H}_{ex-el} = \frac{1}{A}\sum_{\mathbf{p}_{1},\mathbf{p}_{2},\mathbf{p}%
_{1}^{\prime},\mathbf{p}_{2}^{\prime}} W_{eff}(|\mathbf{p}_{1} - \mathbf{p}%
_{1}^{\prime}|,D) X_{\mathbf{p}_{1}}X_{\mathbf{p}_{1}^{\prime}} \hat{l}_{%
\mathbf{p}_{1}^{\prime}}^{\dagger}\hat{c}_{\mathbf{p}_{2}^{\prime}}^{\dagger}%
\hat{c}_{\mathbf{p}_{2}} \hat{l}_{\mathbf{p}_{1}} \ .
\end{eqnarray}
%-------------------------------------------------------------------------

Now let us consider the the two-layer polariton-electron system in
the presence of the polariton superfluidity at $k_{B}T < \hbar
\Omega_{R}$. At temperatures about the Rabi splitting $k_{B}T\gtrsim
\hbar \Omega_{R}$, the upper polaritons states become filled, and
the lower polaritons systems is replaced by the system of the upper
polaritons, which are primarily excitons.

In the presence of the superfluidity we can obtain the Hamiltonian
for the interaction of of quasiparticle excitations in a system of
spatially separated polaritons and electrons by using the Bogoliubov
unitary transformations \cite{Abrikosov}:
%-------------------------------------------------------------------------
\begin{eqnarray}  \label{bog_uv}
\hat{l}_{\mathbf{p}} &=& u_{p}\hat{b}_{\mathbf{p}} + v_{p}\hat{b}_{-\mathbf{p%
}}^{\dagger} \ , \hspace{3cm} \hat{l}_{\mathbf{p}}^{\dagger} = u_{p}\hat{b}_{%
\mathbf{p}}^{\dagger} + v_{p}\hat{b}_{-\mathbf{p}} \ ,  \nonumber \\
u_{p} &=& \left(1 - F_{p}^{2}\right)^{1/2} \ , \hspace{3cm} v_{p} =
F_{p}
u_{p} \ ,  \nonumber \\
F_{p} &=& (\varepsilon_{1}(p) - \xi(p))/\mu \ , \hspace{3cm}
\varepsilon_{1}(p) = \left(\xi^{2}(p) - \mu^{2}\right)^{1/2} \ ,
\nonumber
\\
\xi(p) &=& \varepsilon_{0}(p) + \mu \ ,
\end{eqnarray}
%-------------------------------------------------------------------------
where $\hat{b}_{-\mathbf{p}}^{\dagger}$ and $\hat{b}_{-\mathbf{p}}$
are the
creation and annihilation operators of the quasiparticle excitations, $%
\varepsilon_{1}(p)$ is the energy spectrum of the quasiparticle
excitations in the polariton subsystem, and $\mu = M_{p}c_{s}^{2}$
is the polariton
chemical potential in the Bogoliubov approximation \cite{Berman_L_S}, $%
c_{s}=(U_{eff}^{(0)} n_{p}/M_{p})^{1/2}$ is the sound velocity in
the polariton system, $n_{p}$ is the 2D density of polaritons,
$U_{eff}^{(0)} = 3e^{2}a_{0}/(2\epsilon)$ is the Fourier image of
the polariton-polariton interaction \cite{Berman_L_S}. Substituting
$\hat{l}_{-\mathbf{p}}^{\dagger}$
and $\hat{l}_{-\mathbf{p}}$ from Eq.~(\ref{bog_uv}) into Eq.~(\ref{Ham_exc}%
), we obtain
%-------------------------------------------------------------------------
\begin{eqnarray}  \label{Ham_pol1}
\hat{H}_{ex-el} = \frac{1}{A}\sum_{\mathbf{p}_{1},\mathbf{p}_{2},\mathbf{p}%
_{1}^{\prime},\mathbf{p}_{2}^{\prime}} W_{eff}(|\mathbf{p}_{1} - \mathbf{p}%
_{1}^{\prime}|,D) \sigma (p_{1},p_{1}^{\prime}) \hat{b}_{\mathbf{p}%
_{1}^{\prime}}^{\dagger}\hat{c}_{\mathbf{p}_{2}^{\prime}}^{\dagger}\hat{c}_{%
\mathbf{p}_{2}} \hat{b}_{\mathbf{p}_{1}} \ ,
\end{eqnarray}
%-------------------------------------------------------------------------
where $\sigma (p_{1},p_{1}^{\prime})$ in the presence of the
superfluidity at $T < T_{c}$ is given by
%-------------------------------------------------------------------------
\begin{eqnarray}  \label{sig}
\sigma (p_{1},p_{1}^{\prime}) = \left(u_{p_{1}} u_{p_{1}^{\prime}} +
v_{p_{1}} v_{p_{1}^{\prime}} \right)X_{\mathbf{p}_{1}}X_{\mathbf{p}%
_{1}^{\prime}} \ ,
\end{eqnarray}
%-------------------------------------------------------------------------
and $\sigma (p_{1},p_{1}^{\prime}) = X_{\mathbf{p}_{1}}X_{\mathbf{p}%
_{1}^{\prime}}$ at $T > T_{c}$ without the superfluidity. Let us
mention that at small momenta $\alpha \ll 1$ we have $|X_{p}|^{2}
\approx |C_{p}|^{2} \approx 1/2$.

%====================================================================================

\section{The kinetic equations for distribution function of the
quasiparticles}

\label{ap.ke}

The distribution function of the quasiparticle excitations in the
polariton subsystem $n(\mathbf{p}_{1})$ is represented in a form
%-------------------------------------------------------------------------
\begin{eqnarray}  \label{distex}
n(\mathbf{p}_{1}) = n_{0}(\mathbf{p}_{1}) + n_{0}(\mathbf{p}_{1}) (1 + n_{0}(%
\mathbf{p}_{1})) g_{1}(\mathbf{p}_{1}) \ ,
\end{eqnarray}
%-------------------------------------------------------------------------
where $n_{0}(\mathbf{p}_{1}) =
\left(\exp([\varepsilon_{1}(\mathbf{p}_{1}) -\mu_{qp}(\mathbf{r})]
/(k_{B}T)]) - 1\right)^{-1}$ is the Bose-Einstein distribution
function of the quasiparticles in the polariton subsystem at the
equilibrium, $g_{1}(\mathbf{p}_{1})$ is the contribution to the
quasiparticle distribution function corresponding to the
non-equilibrium
correction due to the gradient of the quasiparticle chemical potential $%
\mu_{qp}(\mathbf{r})$ determined by the external conditions, $k_{B}$
is the Boltzmann constant. We can find $g_{1}(\mathbf{p}_{1})$ by
solving the kinetic equations:
%-------------------------------------------------------------------------
\begin{eqnarray}  \label{kin1}
\frac{\partial n}{\partial\mathbf{r}}\cdot \frac{\partial \varepsilon_{1}(p)%
}{\partial\mathbf{p}} - \frac{\partial n}{\partial\mathbf{p}}\cdot \frac{%
\partial \varepsilon_{1}(p)}{\partial\mathbf{r}} = I_{1}(n) + I_{12}(n,f) \ ,
\end{eqnarray}
%-------------------------------------------------------------------------
%-------------------------------------------------------------------------
\begin{eqnarray}  \label{kin2}
\frac{\partial f}{\partial\mathbf{r}}\cdot \mathbf{v} + \frac{\partial f}{%
\partial\mathbf{p}}\cdot \dot{\mathbf{p}}= I_{2}(f) + I_{21}(f,n) \ ,
\end{eqnarray}
%-------------------------------------------------------------------------
where $I_{1}$ and $I_{2}$ are the collision integrals of the
quasiparticles
in the polariton subsystems and electrons with the impurities, $I_{12}$ and $%
I_{21}$ are the collision  integrals of the quasiparticles with the
electrons, and $f(\mathbf{p}_{2})$ is the electron distribution
function represented in a form
%-------------------------------------------------------------------------
\begin{eqnarray}  \label{distel}
f(\mathbf{p}_{2}) = f_{0}(\mathbf{p}_{2}) + f_{0}(\mathbf{p}_{2}) (1 - f_{0}(%
\mathbf{p}_{2})) g_{2}(\mathbf{p}_{2}) \ ,
\end{eqnarray}
%-------------------------------------------------------------------------
where $f_{0}(\mathbf{p}_{2}) =
\left(\exp((\varepsilon_{2}(\mathbf{p}_{2}) -
\varepsilon_{F})/(k_{B}T)) + 1\right)^{-1}$ is the Fermi-Dirac
electron
distribution function in the equilibrium, $T$ is the temperature, $%
\varepsilon_{F}$ is the electron Fermi energy, $\varepsilon_{2}(\mathbf{p}%
_{2}) = p_{2}^{2}/(2m_{e})$ is the  electron energy spectrum, $g_{2}(\mathbf{%
p}_{2})$ is the contribution to the electron distribution function
corresponding to the non-equilibrium correction  due to the external
electric field $\mathbf{E}$.

We apply the $\tau$ approximation for $I_{1}(n)$ and $I_{2}(f)$:
%-------------------------------------------------------------------------
\begin{eqnarray}  \label{col1}
I_{1}(n) = (n_{0}(\mathbf{p}_{1}) - n(\mathbf{p}_{1}))/\tau_{1}(\mathbf{p}%
_{1}) \ , \ \ \ \ \ \ \ I_{2}(f) = (f_{0}(\mathbf{p}_{2}) - f(\mathbf{p}%
_{2}))/\tau_{2}(\mathbf{p}_{2}) \ ,
\end{eqnarray}
%-------------------------------------------------------------------------
where $\tau_{1}(\mathbf{p}_{1})$ and $\tau_{2}(\mathbf{p}_{2})$ are
the relaxation times of the quasiparticles excitations in the
polariton
subsystem and electrons, respectively. According to Ref.~[%
\onlinecite{Walukiewicz}], $\tau_{2}(\mathbf{p}_{2}) \approx
\bar{\tau}_{2}$ can be approximated by the relaxation time at the
Fermi surface, which can
be determined from the electron mobility $\tilde{\mu}_{e} = e\bar{\tau}%
_{2}/m_{e}$. In a quantum well GaAs/AlGaAs $\tilde{\mu}_{e}$ is
presented in Fig.~2 in Ref.~[\onlinecite{Walukiewicz}].

Since the collision integral $I_{12}$ is only a perturbation with
respect to
$I_{1}$, we neglect it and assume $I_{12} = 0$. The collision integral $%
I_{21}$ has the form \cite{Nikitkov2}
%-------------------------------------------------------------------------
\begin{eqnarray}  \label{i21}
&& I_{21}(g_{2},g_{1}) = 2 \int w(\mathbf{p}_{1},\mathbf{p}_{2};\mathbf{p}%
_{1}^{\prime},\mathbf{p}_{2}^{\prime}) n_{0}(\mathbf{p}_{1}) (1 + n_{0}(%
\mathbf{p}_{1}^{\prime})) f_{0}(\mathbf{p}_{2}) (1 + f_{0}(\mathbf{p}%
_{2}^{\prime}))  \nonumber \\
&& \times (g_{1}(\mathbf{p}_{1}^{\prime}) +
g_{2}(\mathbf{p}_{2}^{\prime}) - g_{1}(\mathbf{p}_{1}) -
g_{2}(\mathbf{p}_{2})) \delta(\varepsilon_{1} (p_{1}) +
\varepsilon_{2} (p_{2}) - \varepsilon_{1} (p_{1}^{\prime}) -
\varepsilon_{2} (p_{2}^{\prime})) \frac{s d^{2}p_{1}^{\prime}}{(2\pi
\hbar)^{2}} \frac{\nu_{e} d^{2}p_{2}^{\prime}}{(2\pi \hbar)^{2}} \ ,
\end{eqnarray}
%-------------------------------------------------------------------------
where $s$ is the level degeneracy (equal to $4$ for excitons in GaAs
quantum
wells), $w(\mathbf{p}_{1},\mathbf{p}_{2};\mathbf{p}_{1}^{\prime},\mathbf{p}%
_{2}^{\prime})$ is the probability of a collision between a
quasiparticle from the polariton subsystem and an electron, which
can be obtained in the Born approximation as
%-------------------------------------------------------------------------
\begin{eqnarray}  \label{born}
w(\mathbf{p}_{1},\mathbf{p}_{2};\mathbf{p}_{1}^{\prime},\mathbf{p}%
_{2}^{\prime}) = \frac{2\pi}{\hbar} \left| W_{eff}(q,D) \sigma
(p_{1},p_{1}^{\prime}) \right|^{2} \ ,
\end{eqnarray}
%-------------------------------------------------------------------------
where $\mathbf{q} = \mathbf{p}_{1}^{\prime} - \mathbf{p}_{1} = \mathbf{p}%
_{2} - \mathbf{p}_{2}^{\prime}$.

Substituting Eq.~(\ref{born}) into Eq.~(\ref{i21}), and expanding
the kinetic equations~(\ref{kin1}) and~(\ref{kin2}) in the first
order with
respect to $\mathbf{\nabla}\mu_{qp}$, we find the functions $g_{1}$ and $%
g_{2}$.

%-------------------------------------------------------------------------
%-------------------------------------------------------------------------

\end{document}